\documentclass[sigconf,authorversion,nonacm]{acmart}

\def\BibTeX{{\rm B\kern-.05em{\sc i\kern-.025em b}\kern-.08emT\kern-.1667em\lower.7ex\hbox{E}\kern-.125emX}}
      
\usepackage{multirow}

\usepackage{amsmath,amssymb,amsfonts}
\usepackage{graphicx}
\usepackage{textcomp}

\usepackage[ruled,vlined]{algorithm2e}
\usepackage{algpseudocode}

\usepackage{xcolor}
\renewcommand{\textcolor}[2]{#2}

\usepackage{braket}

\usepackage{subfigure}

\usepackage[normalem]{ulem} 

\usepackage{amsmath} % for \boldsymbol macro
\usepackage{bm}      % for \bm macro

\floatname{algorithm}{Algorithm}

\begin{document}

% \title{Surpassing Sycamore: Achieving Energetic Superiority Through System-Level Circuit Simulation}
\title{Achieving Energetic Superiority Through System-Level Quantum Circuit Simulation}
%Pioneering Swift Simulation Techniques for Intricate Quantum Circuits}
% \title{Hardware-Efficient Simulation of Quantum Circuits}

\author{Rong Fu$^{1}$ \quad Zhongling Su$^{1}$\quad
 \quad Han-Sen Zhong$^{1,\dagger}$ \quad Xiti Zhao$^{1}$  \quad Jianyang Zhang$^{1}$ \quad Feng Pan$^{2}$ \quad Pan Zhang$^{3}$ \quad Xianhe Zhao$^{2,4,5}$ \quad Ming-Cheng Chen$^{2,4,5}$ \quad Chao-Yang Lu$^{2,4,5,\ddagger}$ \quad Jian-Wei Pan$^{2,4,5}$ \quad Zhiling Pei$^{1,\mathsection}$ \quad Xingcheng Zhang$^{1,\mathparagraph}$ \quad Wanli Ouyang$^{1}$}
\affiliation{$^{1}$Shanghai Artificial Intelligence Laboratory, Shanghai, 200232, China.}
\affiliation{$^{2}$Shanghai Research Center for Quantum Science and CAS Center for
Excellence in Quantum Information and Quantum Physics, University of
Science and Technology of China, Shanghai, 201315, China.}
\affiliation{$^{3}$CAS Key Laboratory for Theoretical Physics, Institute of Theoretical
Physics, Chinese Academy of Sciences, Beijing, 100190, China.}
\affiliation{$^{4}$Hefei National Research Center for Physical Sciences at the Microscale and School of Physical Sciences, University of Science and Technology of
China, Hefei, 230026, China.}
\affiliation{$^{5}$Hefei National Laboratory, University of Science and Technology of
China, Hefei, 230088, China.}
\affiliation{$^{\dagger}$zhonghansen@pjlab.org.cn \quad 
$^{\ddagger}$cylu@ustc.edu.cn \quad 
$^{1,\mathsection}$peizhiling@pjlab.org.cn \quad 
$^{\mathparagraph}$zhangxingcheng@pjlab.org.cn}
\settopmatter{printacmref=false}

\vspace{-0.2cm}
\begin{abstract}
% 背景+问题+解决方法+结果 
%Quantum supremacy -> time - space trade off -> 
Quantum Computational Superiority boasts rapid computation and high energy efficiency. Despite recent advances in classical algorithms aimed at refuting the milestone claim of Google's sycamore, challenges remain in generating uncorrelated samples of random quantum circuits.

In this paper, we present a groundbreaking large-scale system technology that leverages optimization on global, node, and device levels to achieve unprecedented scalability for tensor networks. This enables the handling of large-scale tensor networks with memory capacities reaching tens of terabytes, surpassing memory space constraints on a single node. Our techniques enable accommodating large-scale tensor networks with up to tens of terabytes of memory, reaching up to 2304 GPUs with a peak computing power of \textcolor{red}{561} PFLOPS half-precision.  Notably, \textcolor{red}{we have achieved a time-to-solution of 14.22 seconds with energy consumption of 2.39 kWh which achieved fidelity of 0.002 and our most remarkable result is a time-to-solution of 17.18 seconds, with energy consumption of only 0.29 kWh which achieved a XEB of 0.002 after post-processing,} outperforming Google's quantum processor Sycamore in both speed and energy efficiency, which recorded 600 seconds and 4.3~kWh, respectively.
%We have developed several techniques to overcome the bottleneck of inter-node communication. 
%We have implemented and assessed our techniques, which can handle a tensor network as large as 32TB in just 256 A100 GPUs. 

% Our numerical experiments demonstrate that our approach can achieve higher speed while consuming less enerygy compared to google sycamore.
% The whole computation has cost about xx seconds and consumded xx kwh on a computational cluster with 12*8 GPUs on a 2T task, and about xx seconds and consumded xx kwh on a computational cluster with 32*8 GPUs on a 16T task.

\end{abstract}
\vspace{-0.2cm}

\maketitle
\pagestyle{plain}

\vspace{-0.2cm}
\section{OVERVIEW: A New Era of Speed and Efficiency in Circuit Modeling}
\label{sec:1}

%介绍谷歌量子霸权
% 谷歌的Sycamore处理器在量子计算领域取得了重大突破，它能够迅速完成超级计算机需要数千年才能解决的任务，这一成就推动了量子模拟技术的发展。准确且清晰地传达了Sycamore处理器的重要性和其对量子模拟领域的影响。
%\textcolor{red}{TO DO: Introduction of google sycamore supremacy, need 10,000 years.}
% Within the domain of quantum computing, Google's assertion of "quantum supremacy" with the Sycamore processor constituted a crucial advancement, 
% by generating 1 (3) million uncorrelated samples in 200 (600) seconds that was claimed to require the world's most powerful supercomputer an estimated 10,000 years to perform \cite{Quantum_supremacy}. This assertion has acted as a driving force behind notable progression in the field of classical quantum simulator.
In the field of quantum computing, Google's claim of "quantum supremacy" with its Sycamore quantum processor marks a crucial advancement \cite{Quantum_supremacy}. It generated 3 (1) million uncorrelated samples within 600 (200) seconds, which was claimed to require the world's most powerful supercomputer approximately 10,000 years to complete. This achievement not only propels advancements in quantum computing \cite{jiuzhang1,jiuzhang2,jiuzhang3,zuchongzhi56qubit,zuchongzhi60qubit}, but also serves as a significant catalyst for progress in classical quantum circuit simulation.

% \vspace{-0.1cm}

\begin{figure}[h]
\includegraphics[width=0.9\columnwidth]{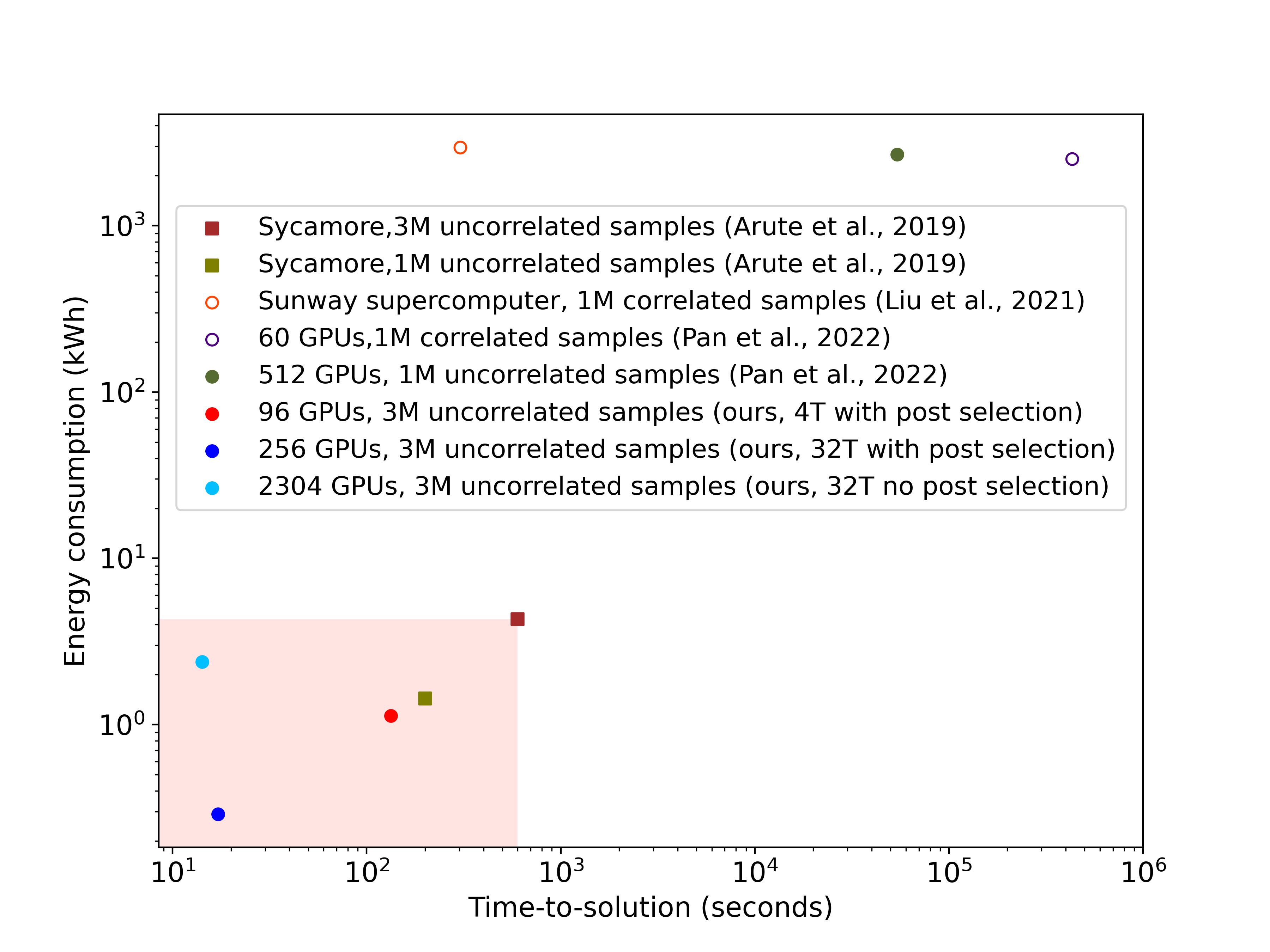}
\caption{Performance of implementations of sampling the Sycamore circuit. The horizontal axis donates the time-to-solution, and the vertical axis donates the energy consumption in the quantum experiment or classical simulations. Circles and squares correspond to classical simulations and quantum experiments, respectively. The hollow circle indicates a correlated sampling loophole in the corresponding classical simulation. The region characterized by misty rose demonstrates superior results in terms of time and energy consumption.}
\label{fig:time_energy_overview}
\end{figure}

%介绍目前的 经典机器：超算和GPU集群 模拟量子计算
% 相关样本，无关样本，时间复杂度、空间复杂度、功耗，算法优劣等
% 2019年以来的论文，按时间排序
% In recent advancements, some exceptionally significant works have considerably enhanced the simulation time for a 53-qubit, 20 cycle Sycamore circuit. This has brought down the original estimated time of 10,000 years as stated in Google's paper to a duration that is now commensurate with quantum experiments. 

% 引出问题：
%，时间、重点在功耗上！多节点传输？还有什么创新点
% 算法不是最sota的？算法优于之前的？需要对比leapfropging，算法的差距，arxiv上的文章，对比；
% 算法上，引用中科院的三篇
% 这篇文章功耗还没有超越的原因是, sub-network 显存受限
% 更小的功耗->更大的sub-nerwork 

%1. 是一个高效的张量网络计算系统。由于显存瓶颈显著影响不同张量网络缩并方案的计算量，通过显存感知、多级并行实现超大模拟的 量子线路模拟/张量网络缩并；
%2. 可以在？*8卡A100上达成更低延时，更优功耗
%\textcolor{red}{TO DO: Introducing the difficulty of RQC simucation.}

% 张量网络的切分

% 用传统的方法来模拟普通的量子电路，所需的计算时间会随着电路的大小成指数增长。grow exponentially with respect to the size of the quantum circuit, 
% 这表明量子电路的复杂性是非常高的
%A critical challenge is constructing a quantum processor that is large enough and accurate to the point where it becomes technically unfeasible for a classical computer to replicate its sampling outcomes. 
%with large-scale and high-fidelity such that its sampling results are intractable for simulation by classical computers. 
The overarching challenge in simulating large-scale quantum circuit is its exponential time complexity, namely, computational cost for quantum circuit simulation increasing exponentially with respect to the size of the quantum circuit \cite{bertels2021quantum, zlokapa2023boundaries, villalonga2020establishing, haner20175, wu2019full}.
% This challenge has been extensively researched \cite{aaronson2016complexity,boixo2018characterizing,bouland2019complexity}.
%thereby rendering such simulations impractical beyond modest circuit sizes.
% 之前有大量的研究致力于将一个大问题分解成许多小的、几乎相同的子任务，这些子任务可以在超级计算机上同时运行。这样做有效地降低了内存需求，从PB减少到TB甚至GB。
To tackle this challenge, a myriad of methods %a significant amount of research 
\cite{Alibaba_19days,Sunway_304s, liu2022validating,guo2019general,li2019quantum}, delineated in Fig.~\ref{fig:time_energy_overview}, have focused on breaking down this large-scale simulation into smaller sub-networks that can be run on a classical super computer, while effectively reducing memory usage from petabytes (PB) to terabytes (TB) or even gigabytes (GB).
However, samples generated in %ref 
\cite{Sunway_304s} exhibit significant correlations among them, thus not considered as faithfully simulating the Sycamore quantum processor. This issue has been addressed in subsequent tensor network researches \cite{512GPUs_15h,60GPUs_5days}. To reduce the computational complexity, \cite{leapfrogging,post-process} have developed a post-processing (aka, post-selection) algorithm, where they selected $k$ bitstrings with the top-$k$ probabilities from $N$ bitstrings. Thereby they enhanced the cross entropy benchmark (XEB) value by a factor of $\ln\left({k}/{N}\right)$. The computational complexity incurred by calculating the probabilities of all samples within any correlated subspace (bitstrings that share some bits) is remarkably low. Post-processing means that computing 3 million independent correlated subspaces, each containing thousands of samples, and subsequently selecting the sample with the highest probability from each correlated subspace. This method not only generates uncorrelated samples but also significantly boosts the XEB value by an order of magnitude.
%Through the application of this approach, they were able to produce uncorrelated samples more rapidly compared to the Sycamore.

% Furthermore, with post-processing technique and  node-level computing approach \cite{leapfrogging}, generating uncorrelated samples faster than Sycamore has been achieved.

% 在处理时间和内存空间之间的权衡。早期研究采用了节点级计算方法，使用单个节点来容纳整个用于收缩计算的子网络，从而最小化节点间通信。这种方法实现了最先进的性能，即86.4秒的处理时间，并将能耗降低到13.7千瓦时，这个数字只比Sycamore的能耗高一个数量级。
% The energy consumption of these works exceeds that of Sycamore by more than three orders of magnitude. In order to balance the trade-off between processing time and memory space, prior research implemented node-level computing approach \cite{Alibaba_19days,leapfrogging} using a single node to accommodate and running one sub-task for contraction, while minimizing inter-node communication.
% This idea achieved a state-of-the-art (SOTA) performance of 86.4 seconds and energy consumption of 13.7 kilowatt hours (kWh) \cite{leapfrogging} which is a hundredfold less than prior studies\cite{Sunway_304s,512GPUs_15h,60GPUs_5days,0.6s}, yet remains one order of magnitude higher than Sycamore's electricity power.

% 最佳收缩方案的计算复杂度与存储空间需求成反比关系。因此，为了进一步降低能耗，不可避免地需要切割更少的索引，这意味着执行更大的子网络以减少计算开销。然而，在考虑节点间通信的带宽时，更大的子网络引入了一个新问题：节点间通信的效率低。
Although tensor network augmented with slicing techniques has demonstrated its advantages in large-scale simulations, %Through analysis, we realized 
the decomposition of quantum circuits into sub-networks typically results in an explosive growth in the computational cost, owing to the heavy overhead from redundant calculations \cite{Lifetime-Based, brandhofer2023optimal, leapfrogging}.
%Under diverse memory constraints defined by varying maximum memory sizes, 

\begin{figure}[h]
\includegraphics[width=0.9\columnwidth]{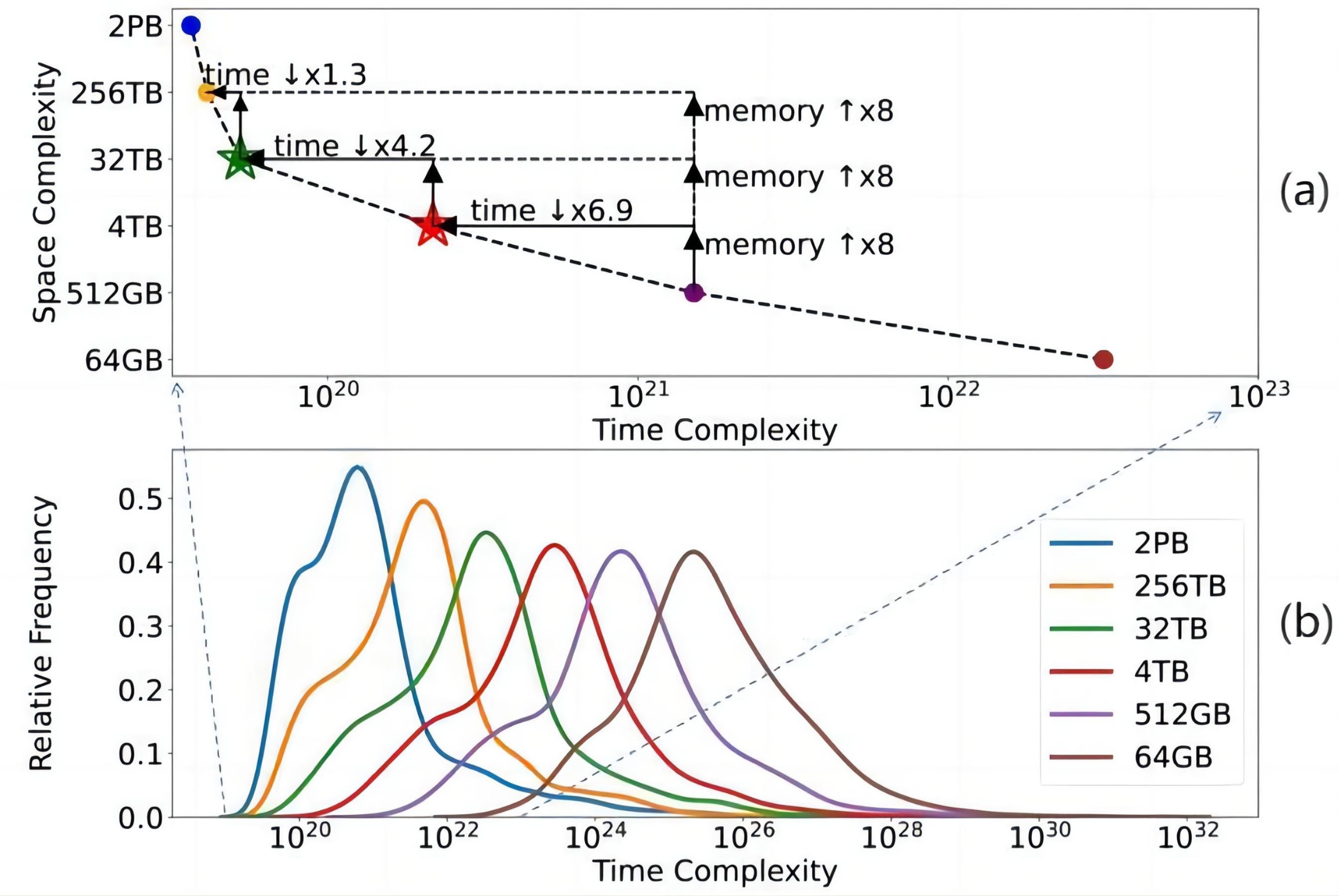}
\caption{The relationship between spatial complexity and temporal complexity. Given a certain amount of memory limits, (a) shows the minimal time complexity of contraction paths where %. under various memory constraint, 
red and green hollow pentagrams represent chosen optimal solutions, \textcolor{red}{4TB and 32 TB}, respectively.
(b) draws time complexity distributions of \textcolor{red}{multiple contraction paths searched by simulated annealing} under various limited memory sizes ranging from 64GB to 2PB.
For each memory constraint, we took its minimum time complexity value in (b) as its optimal contraction path, corresponding to a point with the same color in (a).}
\label{fig:time_space}
\end{figure}

%结论：反比
As shown in Fig.~\ref{fig:time_space}, the computational complexity of the optimal contraction path exhibits an approximate inversely proportional relationship with the maximum memory size.
%Consequently, minimizing energy consumption requires adopting a larger sub-task.
This motivates us to explore the possibilities of minimizing the total subtasks' energy cost through a large-scale distributed-memory systems. However, this approach poses a predicament: when the memory usage exceeds the capacity of a single node, there will be vast %significant 
costs associated with inter-node communication.
The inter-node data exchange not only incurs latency but also severely hinder overall system performance and energy efficiency.

% \begin{figure} [tb]
%         \hspace{-0.3in}
% 	\centering
% 	\subfigure{ 
% 		\label{fig:time_space_probability}
% 		\includegraphics[width=0.55\columnwidth]{1_intro/probability.jpg}
% 	}
% 	%\hspace{-0.3in}
% 	\subfigure{ 
% 		\label{fig:time_space}
% 		\includegraphics[width=0.48\columnwidth]{1_intro/time_space_complexity.jpg}
% 	}
% 	\caption{(a) The time complexity distribution of contraction paths is examined under memory constraints of 64GB, 512GB, 4TB, 32TB, 256TB, 2PB. (b) Relationship between Space Complexity and Time Complexity, where the red and blue hollow pentagrams represent the final solution we adopted.
% 	}	
% \end{figure}

%Therefore, to further decrease energy consumption, it is inevitably necessary to slice fewer indices, which means performing a larger subnetwork to reduce computational overhead. Considering the bandwidth of inter-node communication, the larger subnetwork introduces a new problem: the low efficiency of inter-node communication. 

Facing the aforementioned challenges, we are intended to surpass the capabilities of present-day quantum processors, such as Sycamore, in terms of both speed and power consumption.
%we see the critical need for developing a both quick and energy-efficient classical simulator to outperform modern quantum processor such as Sycamore.
Our major goals are twofold:

\begin{itemize}
\item Firstly, achieving order of magnitude reduction in computation time marks a major breakthrough in the landscape of quantum computing. 
It transcends mere acceleration of the quantum circuit simulation; rather, it represents a fundamental shift in quantum computing where classical computers can not only keep up but also outperform quantum computers even in a computational task, such as random circuit sampling. 
%heralds a transformative epoch poised to redefine quantum supremacy baseline. 
% where quantum computers can evaluate a computational task, random circuit sampling, faster than classical supercomputers. 
% This feat transcends mere acceleration; it embodies a tectonic shift in computational paradigms, suggesting that classical systems are not only viable but also potentially competitive within domains hitherto reserved for quantum processors.
Leading-edge supercomputers not only provide a fast and scalable simulator for increasingly intricate quantum circuits,
but also potentially catalyze the development of wild-range fields, including cryptography and pharmacological research.
This breakthrough could play a pivotal role in fostering environments where the mysteries of the once esoteric quantum realm are unveiled, pushing the frontiers of computational capability.

%Engagement with computational conundrums of exponential complexity, such as those posed by quantum mechanics, prompts a renaissance in algorithmic strategy and computer architecture within the classical supercomputing sector.
%The implications of such a drastic temporal reduction are manifold; it portends a surge in classical algorithm efficiency that may well rival quantum processes in certain scenarios. 
%this marked reduction in computation time necessitates a reimagining of the theoretical boundaries of computation,

%The effects of this time reduction are wide-ranging. It suggests that classical algorithms could become much more efficient, possibly matching the performance of quantum computers in certain situations. This development helps bridge the gap between classical and quantum computing and makes high-performance computing more accessible. It could lead to breakthroughs in various fields, such as cryptanalysis and drug development.

\item Secondly, energy consumption is essential in the realm of quantum computing. Some estimates suggest that quantum technology could reduce energy usage by a factor of 100 to 1000 by processing complex computations much more efficiently. Nonetheless, the total energy efficiency of quantum computing remains an open question, and low-energy solutions requires further exploration.
The quest for energy superiority in circuit simulation, which was once the sole province of quantum processors, can demonstrate the expanding capabilities of supercomputers to solve intricate problems and without relying on an increase in energy.
Furthermore, reducing energy consumption is in concert with global commitments to sustainability and environmental guardianship, fostering the creation of green computing frameworks. In conclusion, an energy-efficient quantum circuit simulator is crucial for ongoing cost-effective analysis of quantum computing and facilitation in driving towards sustainable development in real-world challenges.
% Achieving energetic superiority in circuit simulation, hitherto reserved for quantum processors, greatly prove the capability boundaries of supercomputers of handling complex challenges and expands their market applicability. 
% not only improves cost-efficiency and scalability of quantum simulations, but also has the potential to significantly increase processing power 
% it makes sense to believe that the quantum computer will require less energy than the traditional computer.
% Additionally, reducing energy use aligns with global efforts for sustainability and environmental stewardship, promoting the development of eco-friendly computing paradigms. Therefore, prioritizing low-power, energy-conserving quantum algorithms is essential for the ongoing advancement of quantum computing technologies.
%Increased algorithmic efficiency also enables better scalability for quantum circuits, allowing simulations of larger quantum systems. 

%by engaging with challenges that possess an inherent exponential complexity, like quantum simulation, premier supercomputers have the potential to drive significant algorithmic and architectural breakthroughs within the conventional supercomputing domain. There's wisdom in the ancient Chinese adage from the Classic of Poetry (Shijing), suggesting that "a stone from afar can be used to polish local jade." This implies that the distinct computational paradigms of quantum systems could catalyze a comprehensive reevaluation of classical computing algorithms for various applications.
    
\end{itemize}

Given the above considerations, we proposed the following system-level techniques encompassing algorithmic efficiency, parallel architecture, and precision control:
% 适用于大规模张量网络的三级方案。
% 利用内部和外部资源优化通信效率的新型通信策略。
% 采用混合精度通信的量化方法，强调高效的内核实现，以最小化量化过程中的数据输入/输出（I/O）开销。
% 用于复杂半数据的Einsum扩展简化了高性能库计算，使得einsum操作更加高效。结合自适应缩放方法，可以在最大化计算效率的同时将内存需求减少一半。
% 双缓冲内存管理系统减少了大型张量的分配和释放，从而提高了整体性能和效率。
% 其他优化，例如优化的填充方案，避免了大型分割张量的繁重复制操作，提高了效率。此外，重计算技术在4T子任务中将每个子任务所需的节点数减少了一半，导致计算更加高效。
(1) a three-level scheme tailored for large-scale contraction networks to fully leverage distributed-memory and boost energy-efficiency;
% \item 修改 A three-level parallelization scheme that enables scalability in two
% dimensions: both the size of the tensor network and the number of
% GPUs utilized for executing parallel and independent sub-tasks.
(2) a hybrid communication strategy to maximize intra-node bandwidth utilization and alleviate inter-node data transfer, as well as low-precision quantization to reduce data volume fourfold;
%A hybrid communication strategy to alleviate inter-node communication cost and leverage the high-speed intra-node bandwidth.
%Low-Precision communication to minimize the data size by up to a factor of 4, and a comprehensive study to reveal the impact of various quantization methods.
(3) an efficient Einsum extension for complex-half data, reducing memory requirements by half and leveraging high-speed fp16 tensor core computation. Furthermore, we adopt some additional optimizations, such as an optimized padding scheme that avoids heavy copy operations for large tensors and a recomputation technique which reduces the required number of nodes per sub-task by half in a 4T sub-task, leading to more efficient computing.

% 介绍我们的结果
%Compared to prior high XEB uncorrelated sampling researches, our approach significantly reduces energy consumption by three orders of magnitude. The key lies in our low-complexity algorithm utilizing partial network contraction. We also develop a post-processing algorithm to further enhance XEB values.

%Through analysis, we realized that the optimal contraction scheme's computational complexity is inversely related to storage space requirements. To address this, we developed an advanced, multi-node level parallel tensor contraction algorithm. It increases accessible storage space to a maximum of 32 * 8 × 80 GB = 32 TB for a single sub-network, minimizing computational complexity while maintaining a balanced communication cost.

% 修改 We tested two varying sizes of tensor networks on our three-level parallelism scheme: a 4T and a 32T tensor network, with or without the post-selection technique proposed in \cite{leapfrogging}. This demonstrated close-to-linear scaling using up to 2,304 A100 GPUs.
Our three-level parallelism strategy enables scalability in both the size of the tensor network and the number of GPUs utilized for executing parallel and independent sub-tasks. In our experiments, we tested two varying sizes of tensor networks: \textcolor{red}{a 4TB and a 32TB  tensor network (quantified in the complex-float format)}, with or without the post-processing technique proposed in \cite{leapfrogging}\cite{post-process}. This demonstrated close-to-linear scaling using up to 2,304 A100 GPUs.

\textcolor{red}{We have achieved a time-to-solution of 14.22 seconds with an energy consumption of 2.39 kwh with fidelity of 0.002.} Our top-tier solution comprises a 32T Tensor Network (\textcolor{red}{green} hollow pentagram in Fig.\ref{fig:time_space}) incorporating post-processing technique, which %, when integrated with 256 A100 GPUs, 
impressively cuts the time-to-solution for sampling $3 \times 10^6$ bitstrings to just 17.18 seconds. This represents a significant reduction compared to Sycamore's 600 seconds. Additionally, it consumes a mere 0.29 kWh of energy, markedly lower than Sycamore's energy expenditure of 4.3 kWh \cite{Quantum_supremacy}. 
%With our current understanding, 
Up to our best knowledge, this achievement is the first to demonstrate a clear advantage over Sycamore in terms of energy consumption.
%We remark that to the best of our knowledge, this is the first time that the sampling problem of the Sycamore quantum circuits with n = 53 qubits and m = 20 cycles is solved in minimal energy consumption.

\vspace{-0.2cm}
\section{Current state of the art}
\vspace{-0.2cm}
\subsection{Quantum circuits}

% Similar to their classical counterparts, 
Quantum circuits are constructed from qubits interconnected through a sequence of quantum gates. While classical circuits utilize high and low voltage levels to represent binary states, quantum circuits leverage quantum superposition. The interactions and manipulations of quantum states are dictated by unitary operations known as quantum gates. To acquire information embedded within quantum states, measurements are undertaken. The outcome of a measurement is a collapsed 0–1 bitstring in the chosen measurement basis \cite{pan2023efficient}.
%A qubit is usually described as a linear combination of two orthonormal basis states $\ket{0}$ and $\ket{1}$ (written in the traditional Dirac notation)
It is important to note that measurements annihilate the superposition of quantum states, necessitating the repetition of the entire experiment to gather more insights into the quantum state in question.
Here, we provide a visual illustration of Google's Sycamore \cite{Quantum_supremacy} quantum circuit to clearly showcase the circuit architecture along with quantum gates.

\begin{figure}[h]
\includegraphics[width=0.9\columnwidth]{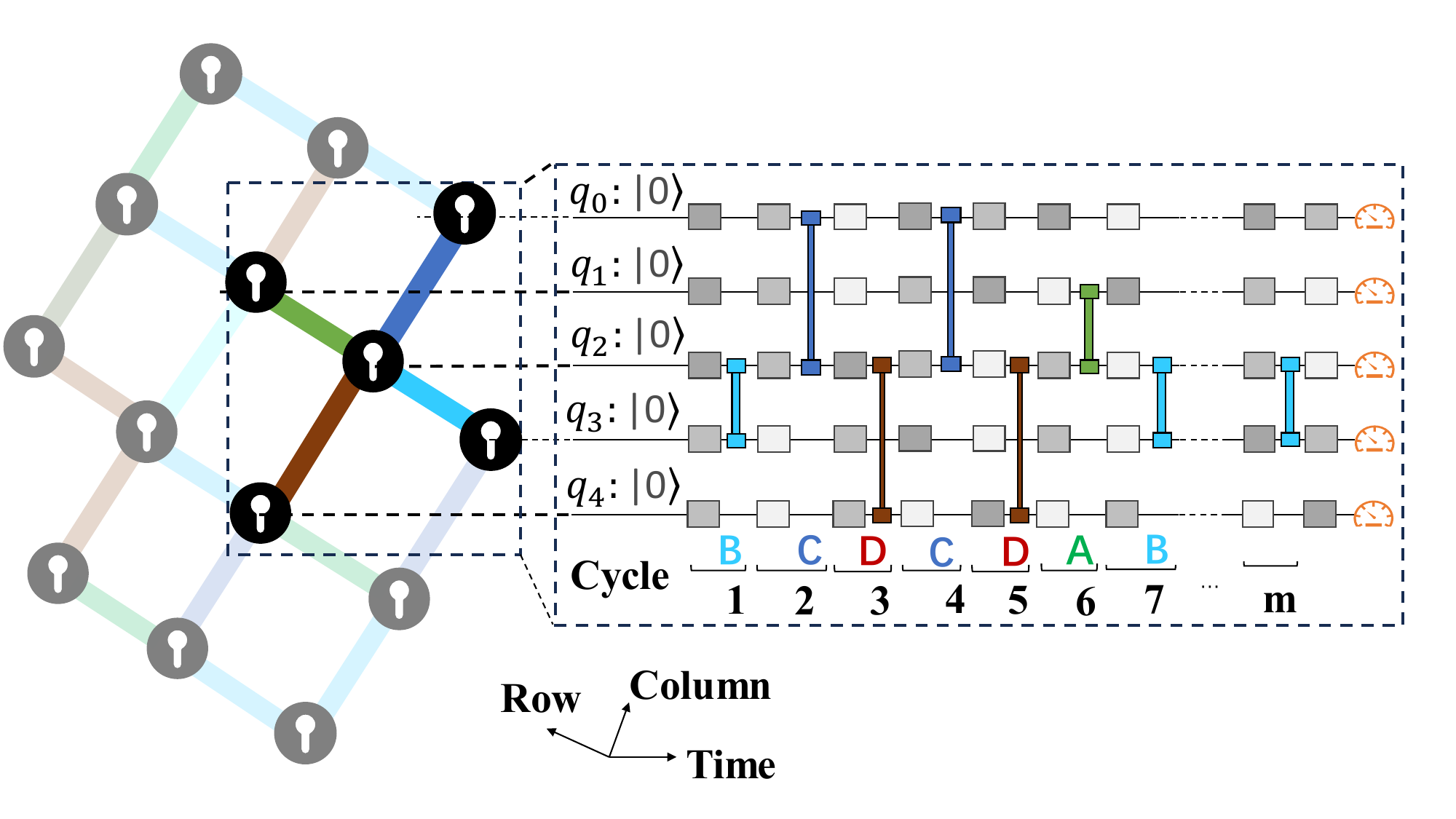}
\caption{Example quantum circuit instance.% used in sycamore's experiment.
}
\label{fig:circuit}
\end{figure}

% \subsubsection{circuit structure}
Fig.~\ref{fig:circuit} provides a portion of a 5-qubit quantum circuit example, within which each random quantum circuits (RQCs) consists of a series of $m$ full cycles followed by a half cycle ending with the measurement of all qubits. Each full cycle involves two steps: First, a single-qubit gate is applied to each qubit. Next, two-qubit gates are applied to pairs of qubits, with different qubit pairs being allowed to interact in different cycles. %Specifically, for supremacy RQCs, they use an eight-cycle sequence (BCDCDAB) which loops through the direct neighbors of every qubit. 
The RQC sequence is repeated in subsequent cycles. The half cycle preceding the measurement exclusively comprises single-qubit gates. 

% \subsubsection{quantum gates}
In Google's Sycamore processor, they configure three distinct single-qubit gates, each of which represents a $\pi/2$-rotation about an axis situated on the equator of the Bloch sphere. Disregarding the global phase, these gates can be represented as follows: {$\sqrt{X}=\frac{1}{\sqrt{2}}\left[\begin{array}{cc}
1 & -i \\
-i & 1
\end{array}\right]$, $\sqrt{Y}=\frac{1}{\sqrt{2}}\left[\begin{array}{cc}
1 & -1 \\
1 & 1
\end{array}\right]$, $\sqrt{W}=\frac{1}{\sqrt{2}}\left[\begin{array}{cc}
1 & -\sqrt{i} \\
\sqrt{-i} & 1
\end{array}\right]$}, and two-qubit gates are determined by the cycle index:
\vspace{-0.2cm}
$$
\mathbf{f S i m}(\theta, \phi)=\left[\begin{array}{cccc}
1 & 0 & 0 & 0 \\
0 & \cos (\theta) & -i \sin (\theta) & 0 \\
0 & -i \sin (\theta) & \cos (\theta) & 0 \\
0 & 0 & 0 & e^{-i \phi}
\end{array}\right],
$$
\vspace{-0.1cm}
where the parameters $\theta$ and $\phi$ of the $\mathbf{f S i m}$ matrix are determined by the qubit pairing. 

% The gate sequence for our pseudo-random quantum circuit generation
% is shown in Fig. 3.One cycle of the algorithm consists of applying single-qubit gates chosen randomly from { X, Y , W} on all qubits,
% followed by two-qubit gates on pairs of qubits. The sequences of gates
% which form the ‘supremacy circuits’ are designed to minimize the circuit
% depth required to create a highly entangled state, which is needed for
% computational complexity and classical hardness.
\vspace{-0.2cm}
\subsection{ RQC simulation methods}

% In general, RQC simulations on classical computers can be classified into two main categories.
RQC simulations on classical computers generally fall into two main categories \cite{Sunway_304s}. 
% Schrödinger algorithm: requires memory space exponential to qubits n; hybrid Schrödinger–Feynman algorithm43: breaks the circuit up into two patches of qubits and efficiently simulates each patch using a Schrödinger method, before connecting them using an approach reminiscent of the Feynman path-integral. Although it is more memory-efficient, the Schrödinger–Feynman algorithm becomes exponentially more computationally expensive with increasing circuit depth owing to the exponential growth of paths with the number of gates connecting the patches
A traditional approach to simulating the evolution of the quantum state is the state vector method \cite{de2019massively}, which tracks the evolution of quantum states using a state vector and time evolution described by the Schrödinger equation. 
%Utilizing an initial state vector and the time evolution described by the Schrödinger equation, one can calculate the final state.
Nevertheless, this would necessitate exponentially large memory space for an $n$-qubit system \cite{vidal2003efficient}, thereby limiting RQC simulations of large numbers of qubits and high entanglement \cite{state_vector}.

An alternative method is to use tensor networks, where an $n$-qubit quantum state can be formulated as a rank-$n$ tensor with $2^n$ complex number entries. Single-qubit and two-qubit quantum gate operations can be represented by rank-$2$ and rank-$4$ tensors, respectively. Consequently, RQC can be converted into tensor networks, which can then be contracted to yield specific final state bitstring amplitudes \cite{villalonga2019flexible, chen2018classical}.

%\subsubsection{state vector simulation}: for qubits <= 50

%\subsubsection{Tensor network simulation}: for qubits > 50

%time-space tradeoff: introducing computation complexity while reducing the space requirement

%slicing technique

%independent and parallel sub-network

%three categories of simulation: single amplitude, full state, and subspace simulation?

Over the past four years, the simulation of Sycamore's RQC has gradually converged into tensor network algorithms. Significant effort has been devoted to two key points: finding the optimal contraction order and streamlining the contraction processes to achieve high-performance calculations. 

% illustration of the tensor network

% 讲一下历史进程
A powerful parallel algorithm for the contraction of tensor networks is stem optimization \cite{Alibaba_19days}, which involves a path of contractions that dominates the overall computational cost. The tensor-based algorithm dynamically decomposes the computational task into numerous smaller subtasks that can be executed in parallel on a cluster. 
Although \cite{Alibaba_19days} demonstrated improvements over conventional methods, these subtasks still involved intense computations within a single node. %are performed on a single node and have high computational cost. % and To avoid inter-node communication, 
\cite{Sunway_304s} introduced an optimized slicing strategy that reduced the space complexity of sliced tensors and alleviated embarrassing parallelism among subtasks.
This strategy, combined with better contraction paths and custom optimizations, allowed for more efficient computing performance than previous methods, such as those provided by the CoTenGra software~\cite{cotengra}. 
%The authors also explore methods for identifying better contraction paths, along with other custom optimizations. As a result, their computing performance outshines that of optimized contraction paths furnished by the CoTenGra software~\cite{cotengra} in the Sycamore simulation. However, using this method requires repeating the contraction process thousands of times to generate uncorrelated samples, an operation that is financially and practically unfavorable.

Pan \textit{et al.} \cite{512GPUs_15h} considered quantum circuits as a $3$-dimensional tensor network, from which they broke edges to form the sub-network. This process can be visualized as drilling holes in the 3D graphical representation of the network. They introduced a method for sparse-state tensor contraction, which allowed for calculating amplitudes of many uncorrelated bitstrings in a single step.
%Based on this, they proposed a sparse-state tensor contraction method, which enables the calculation of amplitudes for a set of uncorrelated bitstrings using a single tensor network contraction. 
This approach is notably more efficient for producing numerous uncorrelated samples than previous techniques. 
% This method is notably more efficient than existing techniques for generating numerous uncorrelated samples with a targeted fidelity.

Furthermore, \cite{leapfrogging} presented a post-processing/post-selection technique where they calculated the probability distribution for each string accurately and obtained top k strings with the largest output probabilities. Through implementing this approach, it was found that it was necessary to execute merely 0.03\% of the total tasks (i.e., contract 0.03\% sub-networks of total networks) out of a total of $2^{24}$ subtasks to reach an XEB value of 0.002. As a result, they significantly improved the cross entropy benchmarking (XEB) values and achieved a groundbreaking faster solution, which has motivated the methodology of this work.

\vspace{-0.2cm}
% \subsection{ Existing RQC simulation experiments}
\subsection{Classical versus quantum: tensor-network-based experiments}
\label{subsec:tensor-network-based}
% parameteres required: Methods? RQC depth and qubits? How many samples? is samples correlated? Fidelity? processor? time? energy consumption? 
Numerous simulations of random quantum circuits have been specifically designed to bridge the quantum-classical simulation divide.
% Researches:
% Zuchongzhi 2:0 (56 qubits)
% The optimal performance of ZuChongzhi 2.0 \cite{zuchongzhi56qubit} can reach up to a system size of 56 qubits and 20 cycles in RQC simulation tasks. Under the premise of XEB being \( (6.62 \pm 0.72) \times 10^{-4} \), ZuChongzhi generated $1.9 \times 10^{7}$ samples and there were low correlated errors among the samples. The estimated time consumption is 1.2 hours.
% qubits：66组成    56效果   20cycles
% time：72分钟
% fidelity（XEB）： (6.62 ± 0.72) × 10^(−4)
% energy：没找到（可能没说）
% samples：1.9 * 10^(7)
% correlated:low correlated errors(原文)
% RQC depth：14（存疑，不太会看）
% Zuchongzhi 2:1 (60 qubit)
Based on tensor network algorithms, ZuChongzhi 2.0 \cite{zuchongzhi56qubit} and Zuchongzhi 2.1 \cite{zuchongzhi60qubit} reached up to a simulation scale of 56-qubits 20-cycles and 60-qubit 24-cycle, respectively. 
Specifically, Zuchongzhi 2.1 required an estimated $1.63 \times 10^{18}$ floating point operations to generate one perfect sample.
% For Zuchongzhi 2.1, the number of floating point operations required to generate one perfect sample is estimated to be $1.63 \times 10^{18}$, as calculated by the python package cotengra \cite{cotengra}.
The random circuit sampling task was completed by Zuchongzhi 2.1 in 4.2 hours and achieved XEB fidelity of $(3.66 \pm 0.345) \times 10^{-4}$. 

%14, may,2020, Alibaba 19 days summit\cite{Alibaba_19days}. By testing the run time of all the basic subtasks, it is estimated that the Sycamore measurement task (20 cycles circuit with a 0.2\% fidelity) can be accomplished within 20 days, using a tensor-based approach and a specific strategy to identify and to optimize the ‘stem’, i.e. the main path for contracting the tensor pairs.
Alibaba presented a tensor network-based classical simulation algorithm \cite{Alibaba_19days}. Through runtime testing of subtasks, it was estimated that the Sycamore task (53 qubits, 20 cycles circuit with a fidelity of 0.2\%) can be accomplished in 19.3 days using the Summit cluster.
% 1 Nov 2021, sunway, \cite{Sunway_2021} 
Xin Liu \textit{et al.} \cite{liu2021redefining} employed contengra to determine a near-optimal tensor contraction order for computational purposes. Consequently, they produced 2,000 perfect samples (or an equivalent of 1 million samples with 0.2\% fidelity) within a span of 6.4 days on the latest generation of Sunway's supercomputer.
% 22 Nov 2021, Sunway, 304 seconds, 1M correlated samples \cite{Sunway_304s}
Yong Liu \textit{et al.} developed an RQC simulator on the advanced Sunway supercomputer. They introduced an optimized path strategy based on tensor network simulation to balance the memory requirements and the number of concurrent computations, resulting in a remarkable achievement of sampling 1 million correlated samples of the Google Sycamore RQC task in just 304 seconds \cite{Sunway_304s}. 

%3 Dec 2021, 0.6 seconds, XEB value of 1.85 × 10−4, failing to meet the benchmark set by Sycamore \cite{0.6s}

Targeting Google's Sycamore RQC problem, Pan \textit{et al.} proposed the big-head approach\cite{60GPUs_5days}, using 60 GPUs for 5 days, they generated $1 \times 10^6$ correlated samples with XEB 0.739, and passed the XEB test. 
%28 Aug 2022, generated $1*10^6$ uncorrelated bitstrings s,fidelity F  0:0037, The whole computation has cost about 15 h on a computational cluster with 512 GPUs 
Another tensor network method \cite{512GPUs_15h} was designed where $1\times10^6$ uncorrelated bitstrings achieved fidelity $F \approx 0.0037$ in a time-cost of 15 hours on a computational cluster with 512 GPUs.
Pan \textit{et al.} also proposed an optimization strategy that transfers the most time-consuming component of the sparse-state tensor network simulation from sparse Einstein summation (einsum) to matrix multiply operations \cite{Alibaba_19days}. 
This optimization was applied to the simulation of 53-qubit, 20-cycle Sycamore circuits, using 2819 A100 GPU hours to verify three million sampled bitstrings.
% For 53-qubit 20-cycle Sycamore circuits with three million sampled bitstrings verification task, the estimated computational resource is 2819 A100 GPU hours.

A recent work \cite{leapfrogging} developed a post-processing algorithm aimed at enhancing XEB values. This algorithm, when implemented with 1,432 NVIDIA A100 GPUs, was capable of yielding 3 million uncorrelated samples with XEB values of $2 \times 10^{-3}$ in a duration of 86.4 seconds, while consuming 13.7 kWh of electricity. 
% \if false
To further trade between spatial complexity and temporal complexity, we extend the work \cite{leapfrogging} through examining the contraction path's time complexity under predetermined memory limits.
While Fig.~\ref{fig:time_space} (b) presents the frequency distribution of all the contraction paths' time complexities under memory sizes from 64GB to 2PB, each point in Fig.~\ref{fig:time_space} (a) illustrates the optimal contraction path with the least time complexity within a certain memory constraints.
%the link between memory capacity and time complexity.
%reveals that increasing memory eightfold cuts the time complexity of the optimal path in half, underscoring the significant role memory plays in optimizing tensor network contractions.
%Within each memory boundary, the lowest value of temporal complexity drawn in Fig.~\ref{fig:time_space} (b) is extracted and designated as the optimal contraction path for that category, which is represented as discrete points in Fig.~\ref{fig:time_space} (a), thereby providing a visual correlation between the memory constraints and their associated temporal complexities.
As the available memory increases by a factor of $8$, the time complexity of the optimal contraction path initially decreases rapidly and then gradually diminishes, eventually converging beyond 32TB memory limits.
This observation highlights the potential of harnessing more memory resources for faster computing.

% \fi

These RQC simulation described above has made significant progress in estimated or actual computation time, %and energy consumption have improved significantly, 
where the gap with Sycamore is no longer substantial.
Nonetheless, even with the state-of-the-art approach, there remains an order of magnitude energy efficiency gap between classical computers and quantum computers in the sampling of RQC tasks.
As \cite{leapfrogging} has solidified the outstanding efficiency of tensor network calculations, our subsequent sections will generate uncorrelated samples based on the optimal solutions for 4TB and 32TB tensor network (marked as hollow pentagrams with red and blue color in Fig.~\ref{fig:time_space} (a)). %algorithm proposed by \cite{leapfrogging,512GPUs_15h}.

% 定义一个新的命令来表示变量
\newcommand{\var}[1]{\textit{\textbf{#1}}}

\vspace{-0.3cm}
\section{Innovations}

The optimal contraction path searching and edge splitting algorithms for the tensor network are based on \cite{512GPUs_15h}. In these methodologies, we employ the technique of drilling holes or breaking edges in the original 3D tensor network to decrease the complexity of a sub-task.
As explained in Subsection~\ref{subsec:tensor-network-based}, leveraging spatial capacity of tensor networks can greatly reduce the computational complexity, for example 4TB and 32TB tensor network. %dividing the tensor network into smaller sub-tasks increases the overall computational complexity.
However, scaling up the tensor network introduces communication overhead, posing a new bottleneck in inter-node communication within the GPU cluster.

To address this challenge, this section proposes various solutions to mitigate inter-node communication issues and harness the potential of the GPU cluster through a three-level parallel scheme embedded with hybrid communication algorithm. Furthermore, we optimize computing resources by converting complex single-precision calculations into complex half-precision calculations. Additionally, we present tailored remedies for specific scenarios encountered in our application domain.
\vspace{-0.2cm}
\subsection{Three-level parallelization scheme}
\begin{figure*}[t]
\includegraphics[width=2\columnwidth]{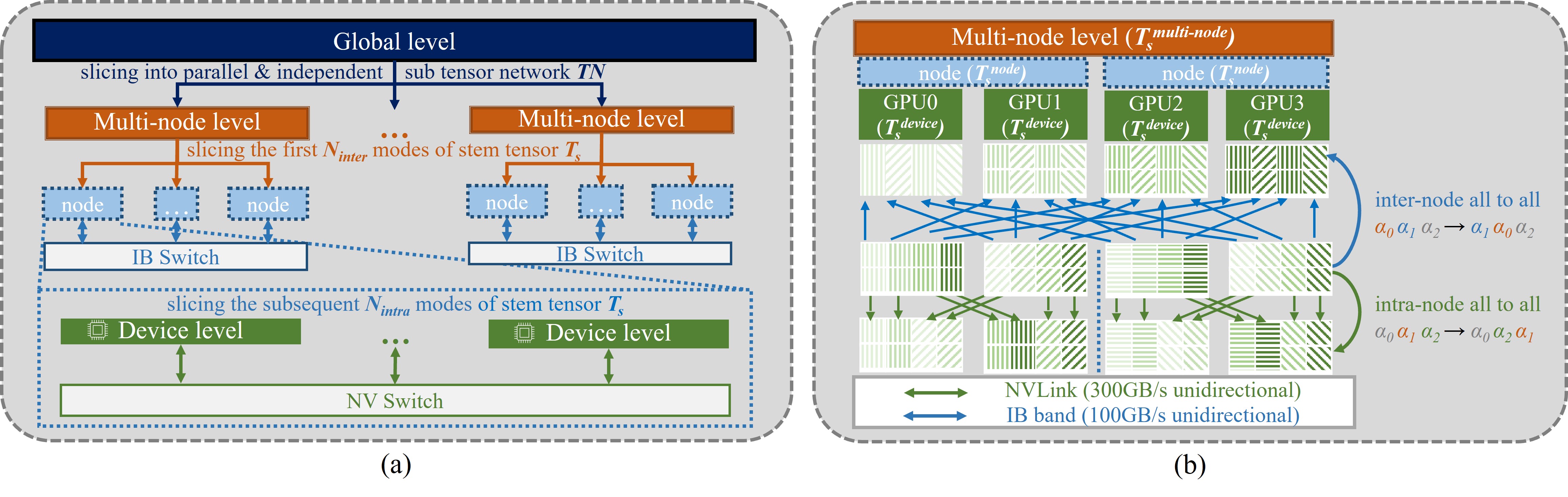}
\caption{Architectural Overview and Example of Parallel Scheme. (a) Overview of the three-level parallel scheme: the task commences at the global level, then the tensor network is partitioned into parallel, independent sub-networks. %, as discussed in \cite{512GPUs_15h, leapfrogging}. 
Data is subsequently segmented across nodes within a multi-node level, interconnected through InfiniBand. Finally, the data is divided into sizes compatible with individual devices  within each node, connected via high-bandwidth NVLink. (b) Example of 2-Node-4-Device Communication: we exhibit a subtask that encompasses two nodes, each hosting two devices, and demonstrate the data permutation occurring across both inter-node and intra-node levels.}
\label{fig:scheme}
\end{figure*}
We categorize our architecture into three hierarchical levels as depicted in Fig.~\ref{fig:scheme} (a): global level, multi-node level, and device level.

Consistent with the approaches described by \cite{512GPUs_15h}, after identifying a near-optimal path, we initially dissect the contraction of the overall tensor network into independent \textcolor{red}{sub-networks} $\bm{TN}$ (eg., 4T $\bm{TN}$ and 32T $\bm{TN}$), which can be executed on separate multi-nodes, labeled as the multi-node level.
Each sub-task, corresponding to the contraction of an independently sliced tensor, is allocated to the multi-node level (indicated by the solid orange rectangle in Fig. \ref{fig:scheme}). Herein, each multi-node level task encompasses the entire sub-network contraction process and is distributed across multiple interconnected nodes via an InfiniBand (IB) network.

\textcolor{red}{In the context of the tensor network, the term "stem path" refers to a sequence of expensive nodes that dominate the overall computation and memory cost \cite{Alibaba_19days}, where "stem tensor" is the tensor associated with the stem path.}

Given a rank-n stem tensor $T_s^{multi-node}(\alpha_0,\alpha_1,...,\alpha_n)$, each of whose rank having a dimension of 2, the first $N_{inter}$ ranks (also referred to as modes) signifies further division of the tensor into $2^{N_{inter}}$ segments. After sliced into a single node, the stem tensor converts to $T_s^{node}(\alpha_{N_{inter}},...,\alpha_{N_{inter}+N_{intra}},...,\alpha_n)$.

The final level of parallelization among disparate devices, exemplified by the dashed green rectangle in Fig.~\ref{fig:scheme} involves specific matrix multiplication and index permutation operations executed on each GPU, interconnected through a high-bandwidth NVLink. Here, stem tensor is further segmented into a reasonable size that can fit into a single device by trimming the subsequent $N_{intra}$ modes, thereby yielding $T_s^{device}(\alpha_{N_{inter}+N_{intra}},...,\alpha_n)$.

% Specifically, the contraction of the entire tensor network occurs at the global level. The tensor network is divided into multiple sub-networks using the slicing method \cite{Lifetime-Based}, which involves fixing the indices over certain edges for each sub-network, enabling each sub-network to be contracted independently. This reduces the space complexity for each contraction while increasing the total time complexity (we will demonstrate this in the evaluation section). We only contract a fraction of the sub-networks, which is performed at the multi-node level. The approximate simulation is similar to the idea in previous work \cite{Solving_the_sampling}. The device node is the smallest unit to perform permutations and contractions.

% \subsection{Hybrid Communication Algorithm}
% \label{section:hybrid_comminication}
Before launching the final network, we introduce a pre-processing step, which involves a hybrid communication algorithm that determine $N_{inter}$ and $2^{N_{intra}}$ mode according to the bandwidth of different storage levels. 

As shown in Fig.~\ref{fig:scheme}, in our classical computational setup, GPUs within a node are connected via NVLink, which has a bandwidth of 300 GB/s. Nodes are connected through InfiniBand, which in our case operates at a bandwidth of 100 GB/s. These IB links are shared by 8 GPUs. As a result, inter-node communication is one order of magnitude slower than intra-node communication.

Our approach divides the workload between intra-node and inter-node communication. This approach takes advantage of the higher bandwidth provided by NVLink, thereby optimizing performance overall.
The hybrid communication algorithm that combines both intra-node and inter-node communication for each step is shown in Algo \ref{alg:Hybrid-communication}.

% To reduce inter-node communication, we distribute the tensor nodes of the stem nodes across $2^{N_{inter}}$ nodes based on the first $N_{inter}$ modes. The $2^{N_{intra}}$ GPUs of a node are then distributed according to the subsequent $N_{intra}$ modes.

\vspace{-0.1cm}
\begin{algorithm}
    \SetAlgoNoLine
    \DontPrintSemicolon
    \LinesNumbered
    \caption{Hybrid communication algorithm}
    \label{alg:Hybrid-communication}
    \SetKwInOut{Init}{init}
    \Init{sub-network consists of initial tensors and the optimal contraction path whose stem is $ \{ ein[i] \}_{i \in \mathcal{I}} $ }
    According to storage levels, determine $N_{inter}$ modes and $N_{intra}$ modes\;
    \For{$i$ in $\mathcal{I}$}{
        $currEin$ = $ein[i]$\;
        \eIf{first-$N_{inter}$ modes contracted}{
        Inter-node communication\;
        }
        {
            \uIf{subsequent $N_{intra}$ modes contracted}{
                Intra-node communication\;
            }
        }
        performer computation of $currEin$\;
    }

\end{algorithm}

Fig.~\ref{fig:scheme} (b) illustrates an example of the rearrangement pattern wherein both $N_{intra}$ and $N_{inter}$ are equal to 1. In this scenario, $a_0$ is related to the inter mode, while $a_1$ is associated with the intra mode. % described above.
During a single tensor contraction, no inter-node communication is required if the first-$N_{inter}$ modes remain uncontracted. However, if any of the following $N_{intra}$ modes are contracted, we rearrange the tensor by swapping the next $N_{intra}$ modes with the following $N_{intra}$ and distribute the tensor based on the new $N_{intra}$ modes. When some of the first-$N_{inter}$ modes are contracted, we rearrange the tensor by swapping the first-$N_{inter}$ modes with the next $N_{inter}$ and distribute it accordingly based on the new $N_{inter}$ modes.

% Given that the primary limitation in communication revolves around inter-node communication, which is substantially larger than intra-node communication, our focus is on minimizing the data transfer among nodes overall. In contrast to the Lifetime-based Optimization strategy discussed in \cite{Lifetime-Based}, we opt for a minimum-data approach that emphasizes reducing the quantity of data exchanged during inter-node communication as opposed to concentrating on reducing the number of communications.

% \input{3_INNOVATIONS/memory/memory}
\vspace{-0.2cm}
\subsection{Customized Low-Precision Communication}
\label{subsec:quant com}
%传输时间很大（complex32*2/64）值得优化 9：6，9：22.5，9：45
% During the execution of a subtask, if the outer modes require reduction, we need to perform all-to-all communication to rearrange the tensor. However, all-to-all communication relies on the inter-node bandwidth (e.g., Infini band), which is relatively slow. When we execute the subtask, the communication time accounts for up to 60.0\%, and the energy consumption is 35\%. As a result, the time and energy consumption becomes a bottleneck in multi-node subtasks. By reducing the accuracy of data, we can reduce the amount of data that needs to be communicated in all to all communication processes. This method not only reduces communication time but also reduces energy consumption.

% 原始版本
% During the execution of a subtask, if outer modes require reduction, an all-to-all communication must be performed to rearrange the tensor. However, this communication relies on relatively expensive inter-node bandwidth, such as InfiniBand. For instance, in a 4T tensor network, the communication time can account for up to 60.0\%, while energy consumption is around 35\%. Consequently, time and energy consumption of inter-node communication become bottlenecks in multi-node subtasks. Therefore, in this section, we provide Low-Precision quantization skills for quantization to alleviate the cost of inter-node communication.

When the first-$N_{inter}$ modes are contracted, %If outer modes require reduction, 
we need to perform an all-to-all communication between multiple nodes as shown in Fig.~\ref{fig:scheme}, serving as a dominant factor in both time and energy usage.
For example, in a 4T tensor network, the inter-node communication time accounts for up to 60.0\%, while energy consumption is around 35\%. Consequently, the time and energy consumption of inter-node communication become bottlenecks. Here we provide low-precision quantization skills %for quantization 
to alleviate the cost of inter-node communication.

% \subsubsection{Quantitative calculation method}
%量化计算通用方式 float-half-int8-int4
% Firstly, we separate the real and imaginary parts of complex64 into float32 real numbers. 

\textcolor{red}{In low-precision quantization, utilizing a single scale and zero offset for the whole tensor could result in substantial computational losses. Therefore, quantizers will divide the tensor into multiple groups, and apply quantization to each group, whose tensor is referred to as group tensor.} According to classic tensor quantization methods \cite{Vectorquantization}, we apply the following general formulation for quantization operator $\mathcal{Q}$ applied to the $i$-th group of tensor $\bm{T}$:
% (see Equation\ref{eq:quantcal}):
% \begin{equation}\label{eq:scale}
% \text{scale} = \frac{\text{val}_{\text{max}} - \text{val}_{\text{min}}}{\text{group}_{\text{max}} - \text{group}_{\text{min}}},
% \end{equation}

% \begin{equation}\label{eq:zero}
% \text{zero} = \frac{{(\text{val}_{\text{min}} \times \text{group}_{\text{max}} - \text{val}_{\text{max}} \times \text{group}_{\text{min}})}}{{(\text{group}_{\text{max}} - \text{group}_{\text{min}})}},
% \end{equation}
 \vspace{-0.2cm}
\begin{equation}\label{eq:quantcal}
\mathcal{Q}\left( [\bm{T}]_i \right) 
%\var{T}_{quant} 
= [\bm{T}]_i^{\text{exp}} \times \text{scale} + \text{zero},
\end{equation}
% \vspace{-0.2cm}
where
$\text{scale} = \frac{q_{max} - q_{min}}{
\max \left( [\bm{T}]_i \right) - \min \left( [\bm{T}]_i \right)
}$ is the scale factor, $q_{max}$ and $q_{min}$ are the range of quantization value we map, and $\max \left( [\bm{T}]_i \right)$ and $\min \left( [\bm{T}]_i \right)$ are the maximum and minimum values of the group tensor $[\bm{T}]_i$. %data we want to perform a single linear quantization.
The zero-point is $\text{zero} = \frac{
q_{min} \max \left( [\bm{T}]_i \right) - q_{max} \min \left( [\bm{T}]_i \right)
}
{
\max \left( [\bm{T}]_i \right) - \min \left( [\bm{T}]_i \right)
}$, and
$\text{exp}$ is an optional parameter to perform \textcolor{red}{exponential} non-linear operation. % and scale and zero are the slopes and intercepts of our linear quantization mapping. 
%The scale and zero have four parameters(see Equation\ref{eq:scale}, see Equation\ref{eq:zero}): 
% --int8 - int4-线性非线性-带入上述公式- group- 双tensor - group与性能规律

 Based on the above equation, we provide three types of quantization: float2half, float2int8, and float2int4 (see Table \ref{tab:quanttable}). The table displays the optimal quantization parameters for each type.
 %Each quantization data type has its range, directly representing its quantization ability. In int4, the fidelity loss of our linear quantization is slightly lower than that of linear quantization (exp=1), while in int8, the opposite is true (exp=0.2). 
 For half and int8 quantization, we compute a global scale and zero-point factor for the entire tensor.
 In contrast, int4 quantization requires a more granular quantization within each group of the tensor rather than the entire tensor, which significantly minimizes fidelity loss \cite{GDRQ}. %leading to dynamic scales and zero-points which also need communication.
 % while in int4, we perform group size data quantization to obtain dynamic scales and zeros that need to be communicated. 
 As different group has different parameters, %which determine its quantization efficiency.
 smaller group sizes result in better fidelity due to tailored scaling and zero-points, but leads to communication overhead.
 % the smaller group size the better the fidelity, but the larger the amount of communication data. 
 Additionally, \textcolor{red}{rounding is incorporated when performing int-quantization.}

 % We use the half and int8 functions to quantify the entire tensor. The $\text{val}_{\text{max}}$ and $\text{val}_{\text{min}}$ parameters of the half and int8 quantization method are ±127 and xxx(half range), respectively. The $\text{group}_{\text{max}}$ and $\text{group}_{\text{min}}$ parameters are the maximum and minimum values of the entire tensor, respectively. After our experiment, the fidelity loss of nonlinear quantization with exp not 1 is relatively small. 
 % \vspace{-0.1cm}
\begin{table}[h]
    \centering
    \caption{Refined quantitative parameters.% for three types of quantization. %Float represents the data type before quantization, while half, int8, and int4 represent the data type after quantization.
    }
    \begin{tabular}{lllll}
        \toprule
        Type & Range &  Exp & Group & Round \\
        \midrule
        float & $\pm 3.4 \times 10^{38}$ &  -- & -- & --\\
        float2half & $\pm6.65\times10^{4}$ &  1 & entire tensor & false\\
        float2int8 & $-128 \sim 127$ & 0.2 & entire tensor & true\\
        float2int4 & $0 \sim 15$ & 1 &  group tensor & true\\
        \bottomrule
    \end{tabular}
     \label{tab:quanttable}
\end{table}

% 使用group原因，动态缩放因子。
% \subsubsection{Implementation Details}
% We use the int4 function to quantify group size data, which has some reason \cite{GDRQ}. Performing quantization once in the group size data instead of once in the entire tensor can greatly reduce fidelity loss caused by low precision of int4.

%The $\text{group}_{\text{max}}$ and $\text{group}_{\text{min}}$ parameters are the maximum and minimum values of the selected group size half data, respectively. 
% The dynamic range provided by the int4 bit type is not sufficient to store the results of all operations simultaneously. The effective single loss scaling factor strategy for int8 becomes infeasible in int4 bit quantization and instead requires the use of different scaling factors for each int4 bit tensor. If the maximum value exceeds the preset threshold, it may be necessary to reduce the scaling factor to avoid overflow; On the contrary, if the minimum value continues to fall below the threshold, it may be necessary to increase the scaling factor to alleviate the underflow problem. The update of the scaling factor needs to be carried out in each step to ensure that the gradient value always remains within the effective representation range of 4 bits.
% Int4 quantization has different group parameters, which determine its quantization efficiency. It is obvious that the smaller the group size, the better the fidelity, but the larger the amount of communication data. 

% \subsubsection{Kernel function design method}
%不使用torch改用kernel
We further enhance performance by crafting custom kernels for all the quantization type above. Dealing with complex-value data, we optimize the kernels through vectorized memory access and fine-tuning for achieving maximum bandwidth utilization and reducing latency. % vectorized memory access and achieved a balanced compute-to-memory access ratio. We selected a kernel parameter of $16$ data points per thread, aligning with the GPU's architecture for peak efficiency.
% Our CUDA kernel, a memory-bound operator, boasts a bandwidth of up to 268GB/s, surpassing the Torch API capabilities.
%
\textcolor{red}{Finally, by adopting int4 quantization kernel, the communication time is decreased by over 85\%  with minimal fidelity loss (compared with float), proving an advantage for various cases.}
% Finally, by adopting int4 quantization kernel, the communication time is decreased by over 10\% with minimal fidelity loss, proving the advantageous for various cases.

\if false
Because Torch does not provide the int4 data type and Torch API does not support kernel fusion, resulting in poor performance. Therefore, we develop our own cuda kernel fusion function for quantization.
%cuda设计思路（向量化，计算访存比）
We use the CUDA kernel to quantify after converting complex data into real data and optimize the kernel function speed by vector memory access and adjusting the computational memory access ratio within the kernel function.
%核函数参数16
We use aligned storage to achieve vector memory access while adjusting the computational memory access ratio. We ultimately chose to calculate 16 data for each thread as the optimal performance parameter for the kernel function. This choice enables us to achieve optimal performance as it best fits the hardware design of the GPU.
%HPC指标
Our CUDA kernel function is a memory-bound operator. Our kernel function has a bandwidth of up to 268GB/s. This is much faster than simply using the torch API.
% and a computing power of up to xxxFlops.
%总结推广

Finally, we achieved significant optimization through quantization with acceptable fidelity loss. Compared to not using quantization, the computational process increased the computation of the quantization kernel function but significantly shortened the communication time. Therefore, we can extend this quantization method to tasks at different data scales.
\fi
% \subsection{Einsum extension for complex-half data}
\vspace{-0.2cm}
\subsection{Expanding Einsum Paradigm for Complex-Half Precision}

Benefiting from the absence of uncontrollable noise, complex-half precision can be adopted in classical quantum simulation to minimize the memory demand.
However, complex-half extensions are not directly applicable. We delve into the einsum paradigm and extend it for complex-half precision support.

Assume that there are three tensors $A, B,$ and $C$, whose ranks (referred to as modes in the context of einsum equation) are $N_A, N_B,$ and $N_C$, respectively. Let $N_{reduce}$ be the number of reduction indices. A general einsum equation % in $ \{ ein[i] \}_{i \in \mathcal{I}} $, 
\begin{small}
\begin{equation}\label{eq:ein_equation}
\alpha_1 \alpha_2 \ldots \alpha_{N_A}, \beta_1 \beta_2 \ldots \beta_{N_B}->\gamma_1 \gamma_2 \ldots \gamma_{N_C},
\end{equation}
\end{small}
can be expressed as a superposition as follows:
\begin{small}
\begin{equation}\label{eq:contraction}
C_{\gamma_1 \gamma_2 \ldots \gamma_{N_C}}=\sum_{\delta_1} \sum_{\delta_2} \ldots \sum_{\delta_{N_{reduce}}} A_{\alpha_1 \alpha_2 \ldots \alpha_{N_A}} B_{\beta_1 \beta_2 \ldots \beta_{N_B}},
\end{equation}
\end{small}
where $\delta_1, \delta_2, \ldots, \delta_{N_{reduce}}$ are the indices to be reduced.
It is crucial to recognize that the formula can only be transformed into a pure General Matrix Multiply (GEMM) operation, not a combination of GEMM and reduction, when the following condition is met: $\delta_1, \delta_2, \ldots, \delta_{N_{reduce}} = (\alpha_1, \alpha_2, \ldots, \alpha_{N_A}) \cap (\beta_1, \beta_2, \ldots, \beta_{N_B})$.
Moreover, 
\begin{small}
\begin{equation}\label{eq:remain_idx}
\gamma_1,\ldots, \gamma_{N_C} = (\alpha_1, \ldots, \alpha_{N_A}) \cup (\beta_1,  \ldots, \beta_{N_B}) \backslash (\delta_1,  \ldots, \delta_{N_{reduce}})
\end{equation}
\end{small}
are the remaining indices for both tensors $A$ and $B$.

To optimize space complexity and enhance calculation speed, a straightforward approach could involve swapping single-precision floating-point numbers with half-precision ones.
% To achieve this, we have two available methods: (1) Transforming the einsum computation into GEMM and subsequent reordering. (2) Leveraging existing HPC libraries such as cutensor and cuquantum, which support the einsum equation.
% The first option results in adding two reordering stages for both the input and output data. Besides, not all equations can be converted into General Matrix Multiplication (GEMM), when an uncommon mode necessitates reduction, this consequently requires an additional reduction computation and more memory space.
% We would prefer to follow the second option. 
However, we encountered a lack of support for contracting complex half-precision numbers within high-performance computing libraries. %, including popular ones such as cutensor and cuquantum.
This deficiency persists even in well-known libraries.
Some libraries like PyTorch support complex half-precision calculations by splitting into real and imaginary parts, which are inefficient due to multiple reads/writes and handling discontinuous data. We introduce a new solution to these challenges.

%They carry out four distinct calculations: real-real, real-imaginary, imaginary-real, and imaginary-imaginary (as detailed in Equations \ref{eq:einsum_real} and \ref{eq:einsum_imag} correspondingly), before combining these results. This process necessitates multiple reads and writes of the dis-contiguous input and output data, resulting in inefficiency.

%\begin{equation}\label{eq:einsum_real}
% C_{real} = Einsum(A_{real},B_{real}) - Einsum(A_{imag},B_{imag})
% \end{equation}

%\begin{equation}\label{eq:einsum_imag}
% C_{imag} = Einsum(A_{real},B_{imag}) + Einsum(A_{imag},B_{real})
% \end{equation}

% We observe in eqs. \ref{eq:einsum_real} and \ref{eq:einsum_imag} that there are two inefficiencies: multiple reads of inputs and accessing discontinuous data (real and imaginary numbers). To overcome these issues, we propose a new approach.

% As indicated in Section \ref{sec:memory}, computational processes can be subdivided into three categories based on the nature of the computations being carried out. 

%From a data-volume standpoint, the principal focus of memory usage and computational boundaries lies in the einsum computation involving a large tensor with gigabytes (GB) in size and a smaller tensor measuring in kilobytes (kb) or megabytes (mb). In this situation, one tensor substantially outweighs the other in terms of size.
%This requires careful data management to ensure efficient resource allocation and minimize the impact of large datasets on performance.

To enhance readability, we generally denote $A$ as the larger input, $B$ as the smaller one, and $C$ as the output. In this case, $B$'s IO operation is often negligible compared to that of $A$ and $C$, so our primary focus is on data access for $A$ and $C$.

% The approach is to divide the complex half of the input data into halves, with the last mode indicating whether it represents the real or imaginary part. It is worth noting that this change only multiplies the number of elements by 2, but the size in bytes of inputs and outputs and their memory locations, remain unchanged since a complex half occupies 4 bytes and a real half plus an imaginary half also occupy 4 bytes.

A straightforward attempt is to simply append a mode that indicates whether it represents the real or imaginary part at the end of each component in Equation \ref{eq:ein_equation}, as shown in Equation \ref{eq:eq_modified}. However, this approach is incorrect since $\alpha_{N_{A+1}}$ and $\beta_{N_{B+1}}$ are the indices to be reduced, and $\gamma_{N_{C+1}}$ is the remaining index, we do not have any index generating $\gamma_{N_{C+1}}$ in either of the inputs, which necessitates a modification of Equation \ref{eq:eq_modified}:
\begin{equation}\label{eq:eq_modified}
\alpha_1  \ldots \alpha_{N_A} \alpha_{N_{A+1}}, \beta_1 \ldots \beta_{N_B}  \beta_{N_{B+1}}->\gamma_1  \ldots \gamma_{N_C} \gamma_{N_{C+1}}.
\end{equation}

Since introducing a supplementary mode necessitates duplicating the size in bytes, the most efficient approach is to include an extra mode for tensor B, as it is smaller than tensor A. The final equation becomes:
\begin{equation}\label{eq:eq_modified2}
\alpha_1  \ldots \alpha_{N_A} \alpha_{N_{A+1}}, \gamma_{N_{C+1}} \beta_1 \ldots \beta_{N_B}  \alpha_{N_{A+1}}->\gamma_1  \ldots \gamma_{N_C} \gamma_{N_{C+1}},
\end{equation}
where tensor $B$ is padding from $[B_{(real, imag)}]$ to \\$[B_{(real, -imag)}, B_{(imag, real)}]$. Here, we substitute $\beta_{N_{B+1}}$ with $\alpha_{N_{A+1}}$ because they are identical by definition from Equation \ref{eq:remain_idx}.
%To overcome this hurdle, we propose leveraging the high computational power of GPUs to solve this problem. By transmuting the problem of computing the einsum of complex half-precision numbers into that of computing the einsum of half-precision numbers alone, we could benefit from the performance gains provided by the graphic processing units. 

% The only remaining task is determining the elements in the new tensor B, assuming we simply pad tensor B as $[B_{(real, imag)}] -> [B_{(real, imag)}, B_{(real,imag)}]$. In this case, Equation \ref{eq:eq_modified2} would yield eq.\ref{eq:einsum_real_false} and eq.\ref{eq:einsum_imag_false}, rather than the correct results! Fortunately, the solution is straightforward: alter the padding method to $[B_{(real, imag)}] -> [B_{(real, -imag)}, B_{(imag, real)}]$. This approach produces exactly the same outcomes as eq.\ref{eq:einsum_real} and eq.\ref{eq:einsum_imag}, requiring only one calculation and a minimal amount of padding.

% The only remaining task is determining the elements in the new tensor B, the solution is straightforward: padding tensor B to $[B_{(real, imag)}] -> [B_{(real, -imag)}, B_{(imag, real)}]$.

% \begin{equation}\label{eq:einsum_real_false}
% C_{real} = Einsum(A_{real},B_{real}) + Einsum(A_{imag},B_{imag})
% \end{equation}

% \begin{equation}\label{eq:einsum_imag_false}
% C_{imag} = Einsum(A_{real},B_{real}) + Einsum(A_{imag},B_{imag})
% \end{equation}

Here's an example, where we have $a_1a_2, b_1 \rightarrow a_1b_1$, with $a_1$ being $[[(1 + 2i),(3 + 4i)]]$ and $b_1$ being $[(5 + 6i)]$. This leads to a result of $[(-7 + 16i),(-9 + 38i)]$.
Following the provided method, we transform the corresponding complex tensor $A$ into $[[1, 2],[3,4]]$ and rearrange tensor $B$ to $[[5, -6],[6, 5]]$. By applying the modified equation $a_1a_2,c_0b_1a_2\rightarrow a_1b_1c_0$, we obtain the same exact result, $[[-7, 16]], [[-9, 38]]$.

\vspace{-0.2cm}
\subsection{Optimization for special cases}
We introduce several specialized skills. \textcolor{red}{They are not as generalized as previous introduces methods, but they are useful for certain tasks.}
\vspace{-0.2cm}
\subsubsection{Recomputation} 
\label{recomputation}
In the 4T tensor network, we observe the following features: there are only four steps with values above 1T, one of which is 2T; 2. During and after these four steps, there is no data communication. Based on this information, we designed recomputation techniques. The main idea is that instead of calculating the entire large tensors at a time, we compute half of them each time. Specifically, we begin at the start, just before the generation of the 1T tensor. We store and compute only half of the tensors and proceed until the end. Then, we restart from the middle, read the remaining half, and calculate the second half. Finally, we concatenate the two parts. This technique substantially reduces the necessary nodes by two, concurrently diminishing the data volume for all-to-all communication, considering that $N_{inter}$ is concurrently diminished by 1.
\vspace{-0.2cm}
\subsubsection{Tensor contraction during sparse state}
% 在前文的描述中，sparse state发生在张量网络计算的最终阶段。它本身是不连续且重复的，需要拷贝张量进行计算，然而，因为分配了双buffer的原因，显存不太够且不支持进行大张量计算，所以我们把大张量分成能够放进显存的小片，分的片数由当前的显存剩余容量而定，接着迭代计算每一片张量。
As described in \cite{512GPUs_15h}, sparse state occurs in the final stage of tensor network computation. It is inherently discontinuous and repetitive, requiring the copying of tensors for computations. However, due to the allocation of a double-buffer, the GPU memory is nearly exhausted and cannot support the loading of large tensor calculations, so we divide the larger tensor into smaller chunks that can fit into the current GPU memory, and the number of chunks is determined by the current remaining capacity of the GPU memory. Then we compute each tensor chunk iteratively.

\begin{figure}[h]
\includegraphics[width=0.9\columnwidth]{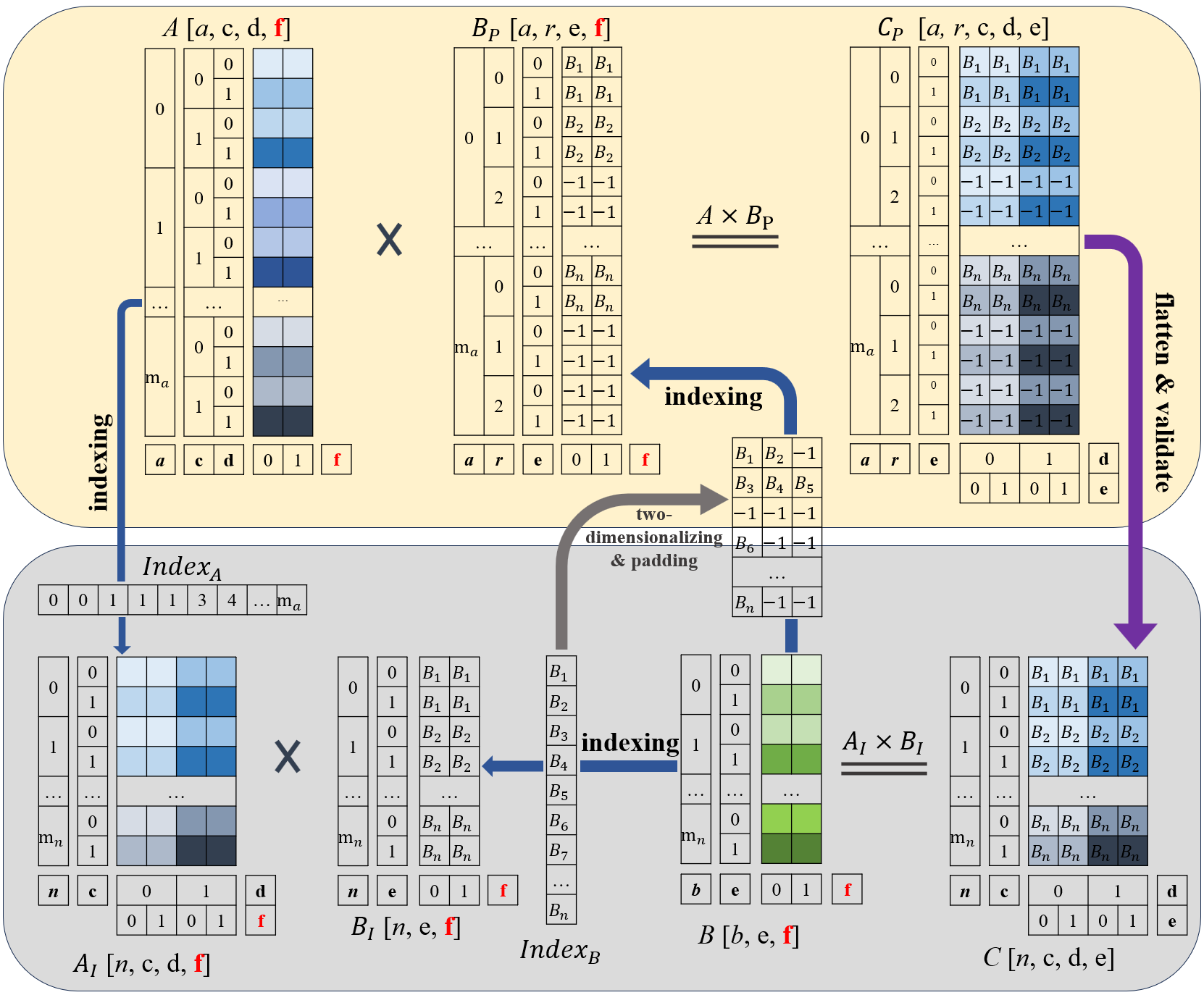}
\caption{The bottom part is tensor multiplication by traditionally retrieving tensors through indices. The top part is multiplication between source tensor and padding-tensor which retrieved through a 2-dimensionalizing index. $A_{I}$, $B_{I}$ are multiplication input tensors which are indexing from source tensors A, B. $B_{P}$ is a padding-tensor. C and $C_{P}$ are different products of two multiplications.}
\label{fig:sparse state}
\end{figure}

% 张量缩并包括维度重排和矩阵乘法这两步。然后，由于英伟达A100的加持，我们得以不再局限于一对张量之间的缩并，利用cuTensor库函数，我们可以实现多对张量同时计算。如图，我们可以在进行矩阵乘法之前通过索引I_A and I_B获得多个指定位置的张量以形成大张量A_I and B_I。我们设计了两种索引方案，以适用于不同的索引分布情况。
% Typically, tensor contraction involves dimension reordering and matrix multiplication. However, in this paper, we implement parallel computation using GPU and CUDA kernel functions. Therefore, we need to add a step before this - obtaining the combined tensor according to the index. As shown in Fig. x, a combined tensor can be obtained from any position of the Element Tensor through an index array as an input tensor. Then, the dimensions to be contracted are reordered to the end. Finally, the two input tensors, dimension ordering, and result storage location are passed into the CUDA function to iteratively calculate the contraction results of each piece.
Tensor contraction involves dimension reordering and matrix multiplication \cite{Sunway_304s}. Then, with the enhancement of the NVIDIA A100, we are no longer limited to the contraction between a pair of tensors. By utilizing the cuTensor library function, we can perform computations on multiple pairs of tensors simultaneously. As shown in Fig.~\ref{fig:sparse state}, we can obtain multiple tensors at specified positions through the indices $Index_A$ and $Index_B$ to form the large tensors $A_I$ and $B_I$ before performing matrix multiplication. We have designed two indexing schemes to accommodate different index distribution scenarios.

%rank
%a dimension size of
Let \textit{A[a,c,d,f]} and \textit{B[b,e,f]} be input tensors, \textit{C[n,c,d,e]} be the contraction product, all with two dimensions except the outermost ranks \textit{a, b, n}, and let \textit{f} be a common rank to be contracted. $m_a$ and $m_b$ respectively are chunk-size of \textit{A} and \textit{B}, so the $m_n$ is the same of \textit{C}. Usually, $Index_A$ and $Index_B$ are arrays with $m_n$ integers from 0 to $m_a$ or $m_b$, then $C=A_I \times B_I$. However, when there is a large amount of repeated data in the \textit{Index} corresponding to input tensors of high-rank, just like $Index_A[0, 0, 1, 1, 1, 3, 4, ..., m_a]$, it will be very expensive in this way. So we use the input tensor \textit{A} directly. In the $Index_B$, $B_i \leq m_b(i = 0, 1, ..., n)$. Thus, for the rightness of retrieving $B_I$, we need padding $Index_B$ to new 2d-index which size is $m_a*m_r$($m_r$ is max repeat count of one number in $Index_A$), and excess positions are replaced with -1. In this case, $m_r$ is 3 since 1 had appeared 3 times. To avoid ambiguity, we call this new $B_I$ as $B_P$, and because \textit{B} as smaller tensor compared with \textit{A}, so it won't increasing too much costs in this step. Finally, we got $C_P=A \times B_P$, then \textit{C} can derive from $C_P$ through flatten its outside ranks \textit{a, c} and then extract valid tensors in it.

\vspace{-0.4cm}
\section{Evaluation}

We begin by presenting our experimental setup. Following that, we discuss the tools utilized in our performance analysis. Next, we conduct an evaluation of low-precision communication and provide usage advice for communication in RQC simulation. We proceed to assess the proposed technique step by step and identify the optimal solution for a single multi-node level subtask. Finally, we present and analyze the results of testing two large-scale tensor networks (specifically, 4T and 32T). % and compare their performance with that of Google's Sycamore.

\vspace{-0.2cm}
\subsection{Experiment setup}

% 为什么用两个集群？

We employ 80GB A100 GPUs, featuring peak FP16 Tensor core performance of 312 TFLOPS. Eight A100 GPUs within each node are interconnected via NVlink, providing a unidirectional speed of 300 GB/s, while nodes are connected via InfiniBand with a unidirectional speed of 100 GB/s.
The task is implemented based on Pytorch framework (version 2.1), einsum calculation is performed using Cutensor (version 1.7.0) with cudatoolkit of version 12.0.  
% We adopt a hybrid communication approach that combines intra-node and inter-node communication approach (see section\ref{section:hybrid_comminication}) for a subtask to mitigate the time and energy cost associated with communication.

% We use the Complex32 as data type in contraction, and int4 as data type in communication.
\vspace{-0.2cm}
\subsection{Measurement tools}
% \subsubsection{Monitor power consumption in real-time using NVML}
% 为了实时监控GPU功耗，我们采用了一种利用子进程持续调用NVML库python接口的方法
For monitoring GPU power consumption in real-time, we have adopted a method that creates a subprocess continuously invoking the NVML library \cite{nvml} with Python interface.
% 这个子进程运行在device level，将它的机器序号作为参数传递给指定函数，这样函数就可以每20ms记录和保存一次每张卡的相对时间和瞬时功率。因为它是通过启动另一个进程实现的，所以它几乎并不会对主进程造成影响
This subprocess runs on the device level, passing its own machine rank to a specific function as a parameter, so the function can capture and store both the relative timestamp and the instantaneous power for each graphics card, performing these tasks at approximately 20-millisecond intervals. Since it is implemented by initiating a separate process, it almost does not impact the performance of the main process.

% 在运行完张量网络后，我们使用积分计算整个过程的总功耗并在global level上进行加和，它能很大程度上精确地反应这个过程的能耗
At the end of the tasks, we used the method of infinitesimal integration to calculate the total energy consumption throughout the process and sum up it on the global level as a result, which can reflect the consumption of the entire operation process. In Table \ref{table:power}, we present the measured power under three distinct scenarios.

\begin{table}[h]

\caption{Measured Power per A100 GPU
% -(1) No operation, (2) Communication, and (3) Computation.
}
\label{table:power}
\begin{tabular}{cc}
\hline
              & Power per A100 GPU \\ \hline
Idle          & 60 W               \\ \hline
Communication & 90$\sim$135W       \\ \hline
Computation   & 220$\sim$450W      \\ \hline
\end{tabular}
\end{table}

% \subsubsection{Time and stoarge measurement}
% Time measurements for the global-level simulation were obtained using a combination of \\$torch.cuda.synchronize()$, $torch.distributed.barrier()$ and system timer.

%解释压缩公式
Based on our designed quantization method in Subsection~\ref{subsec:quant com}, we define a quantization compression rate (CR (\%), %shown in \eqref{eq:compress}
), which denotes the percentage of data to be compressed for communication, 
\begin{small}
\begin{equation}\label{eq:compress}
\text{CR(\%)} = \frac{{\text{sizeof}(T_{\text{scales}}) + \text{sizeof}(T_{\text{zeros}}) + \text{sizeof}(T_{\text{quant}})}}{{\text{sizeof}(T_{\text{origin}})}},
\end{equation}
\end{small}
where $T$ represents the tensor.

% \subsubsection{Fidelity measurement}
The results are compared to the benchmarks using the similarity function presented below, with similarity denoted as fidelity:
\begin{small}
\begin{equation}\label{eq:fidelity}
fidelity = \left|\frac{\sqrt{\langle benchmark, result \rangle^2}}{\lVert benchmark \rVert  \lVert result \rVert}\right|^2.
\end{equation}
\end{small}
\vspace{-0.3cm}
\subsection{Assessment for low-precision communication}
%实验规模指标
% We conducted experiments on a small data scale (using a total of 16 A100 cards from 2 nodes with inter-node bandwidth of 100GB/s and nvlink used intra-node bandwidth to run the single multi-node level task) and divided the quantified fidelity by the complex64 fidelity as relative fidelity. In the parameter selection of the quantitative communication experiment, We have comprehensively considered the relative fidelity loss and the performance improvement on time and energy consumption.

In this section, we explore the most effective method of data quantization for communication with the dual goals of minimizing energy consumption and achieving high-fidelity results. % that closely resemble the original single-floating point data.
First, we analyze precision sensitivity within the stem path of the 4T tensor network. % with 8 nodes. 
%Following this, 
We then evaluate the impact of quantization for both intra-node and inter-node communication, and identify the quantization strategy for tackling the fidelity-energy trade-off. %identify the optimal quantization method that balances energy conservation and data fidelity.examining the impact on both energy and time, and find it results in negative returns.

% We then explore inter-node communication, conducting end-to-end experiments on a 4T tensor network with 8 nodes to identify the optimal quantization method that balances energy conservation and data fidelity.

\vspace{-0.2cm}
\subsubsection{Single step of quantization}
\label{sec:single_step_quant}
\begin{figure}[h]
\includegraphics[width=0.9\columnwidth]{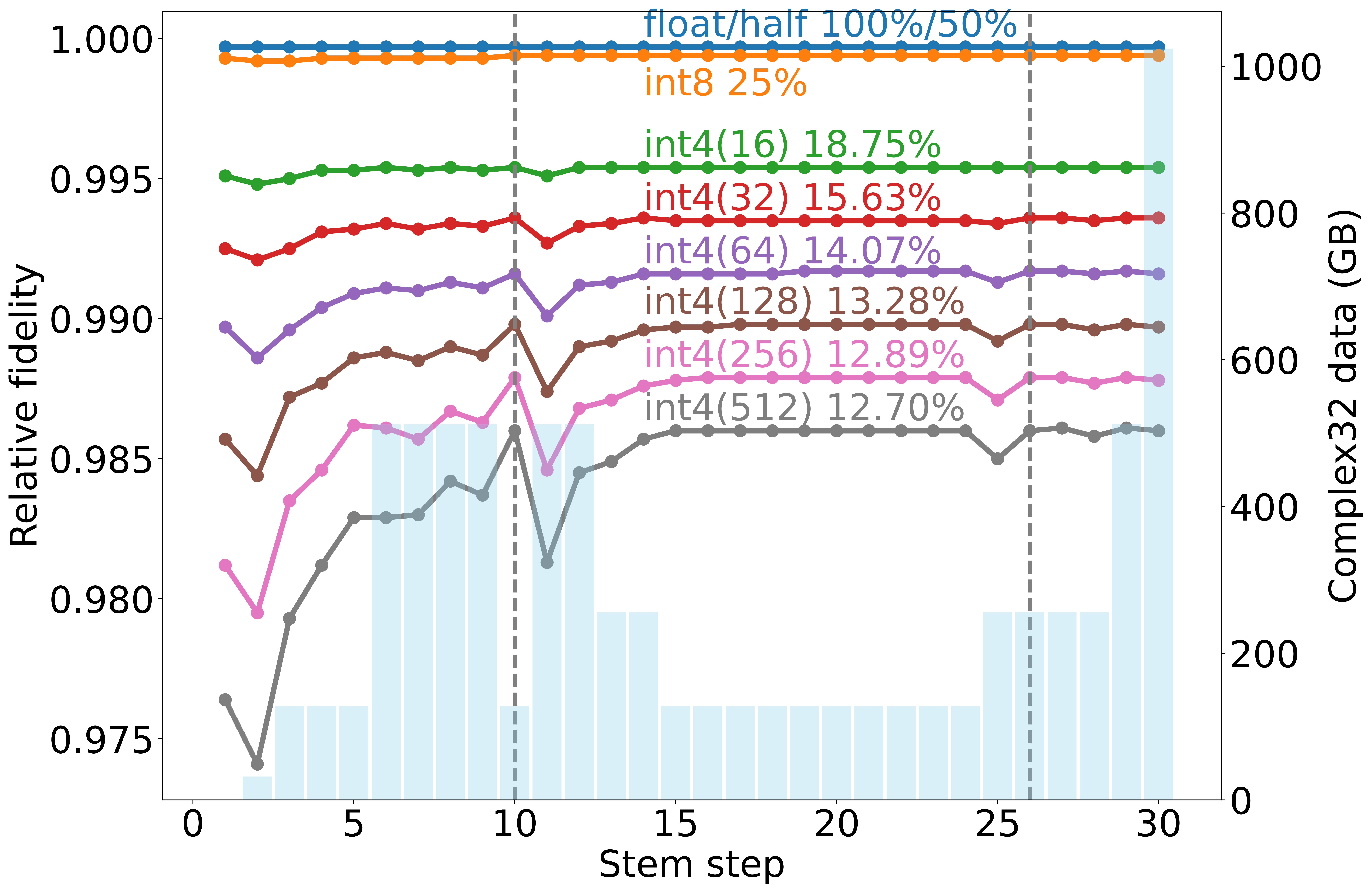}
\caption{
Quantitative compression rate (see Equation\ref{eq:compress}) and relative fidelity in single step quantization. The curve represents the relative fidelity, calculated as the quantified fidelity divided by the complex64 fidelity. The percentage number represents CR (\%). 
% The textual description of curves is the quantization method and their quantization compression rate. 
% The dotted line represents the communication location we have chosen.
%The marked dashed line on the graph is the quantization steps we ultimately chose.
}
\label{fig:quantize1}
\end{figure}

%乘积关系
% It is obvious that the quantization communication process requires multiple quantization steps, and the effect of multi-step quantization is the product of the corresponding single step quantization effect.
We conducted stepwise quantization experiments based on single-step quantization %and finally determined the optimal communication step 
(see Fig.~\ref{fig:quantize1}).
%step图片
% 解释前期保真度低
%图片规则
Relative fidelity is less stable and has lower performance in the early stage of the task. The reason is that quantifying in the early stages of the task will result in more error accumulation than quantifying in the later stages of the task. Furthermore, the relative fidelity is independent of the amount of communicated data. 
%传输step选择
Therefore, we attempt to quantify in the later stages of the task and choose a larger amount of data for quantification as much as possible to achieve higher returns. The marked dashed line on the graph is the quantization steps we ultimately chose.
\subsubsection{Assessment for quantization of intra-node communication}
%总起：开始传输实验
% After determining the communication step, we use different quantization methods to run the single multi-node-level task to compare their performance. In the end, we obtained the quantization method that maximizes performance improvement considering relative fidelity loss as the scheme for our entire experiment.

The time of all-to-all communication is
\begin{small}
\begin{equation}\label{eq:time_all2all}
 T_{all2all} = \frac{Data\ Amount}{bandwidth} * \frac{N}{N-1} * \frac{1}{r},
\end{equation}
\end{small}
where $N$ represents the number of devices within a node, which is 8 in this case, and r denotes the bandwidth utilization rate in all-to-all communication, which is approximately 50\% in practice.
The energy consumption is proportional to:
\begin{equation}\label{eq:energy}
energy \propto \alpha*T_{all2all} + \beta*T_{calculation},
\end{equation}
where \( \alpha \) and \( \beta \) are the coefficients of power consumption. Empirical data shows that the ratio \( \frac{\alpha}{\beta} \) is approximately equal to \( \frac{1}{3} \). 
For every 1GB of data, the quantization kernel takes 4.25ms to process, whereas the reduction in communication time resulting from quantization, as per equation \ref{eq:time_all2all}, is a mere 4.78ms. This suggests that the time-saving achieved through quantization is, indeed, insignificant.

Upon further analysis, taking into consideration Equation \ref{eq:energy}, the increase in overall energy consumption, denoted by the inequality \(-\alpha*4.78 + \beta*4.25 > 0\), results in a decline in performance. This revelation, combined with the earlier mentioned reduction in fidelity as described in Section \ref{sec:single_step_quant}, conclusively demonstrates that implementing quantization techniques within intra-node communication yields negative outcomes. As a result, the utilization of quantization for intra-node communication is not considered advantageous, and would not be adopted in the final task.

%节点内理论数据 负收益

% We use nvlink for intra-node communication, as the bandwidth of nvlink is high, the saved communication time and increased kernel time are not significant compared to each other. Another point is that at the same time, the communication energy consumption is only 1/2 to 1/4 of the calculated energy consumption, so overall, the energy consumption of the task has increased. However, quantization will inevitably lead to a decrease in fidelity. From a data perspective, when conducting experiments using int4(128). For every 1GB of data, the kernel function takes 4.25ms to process, while according to the initial communication time, the quantified communication time is only reduced by 4.78ms. However, the total energy consumption has increased, resulting in a decrease in performance. Therefore, quantization between intra-node communication brings negative benefits, so we do not use it.
\subsubsection{Assessment for quantization of inter-node communication}
\begin{figure}
\includegraphics[width=0.9\columnwidth]{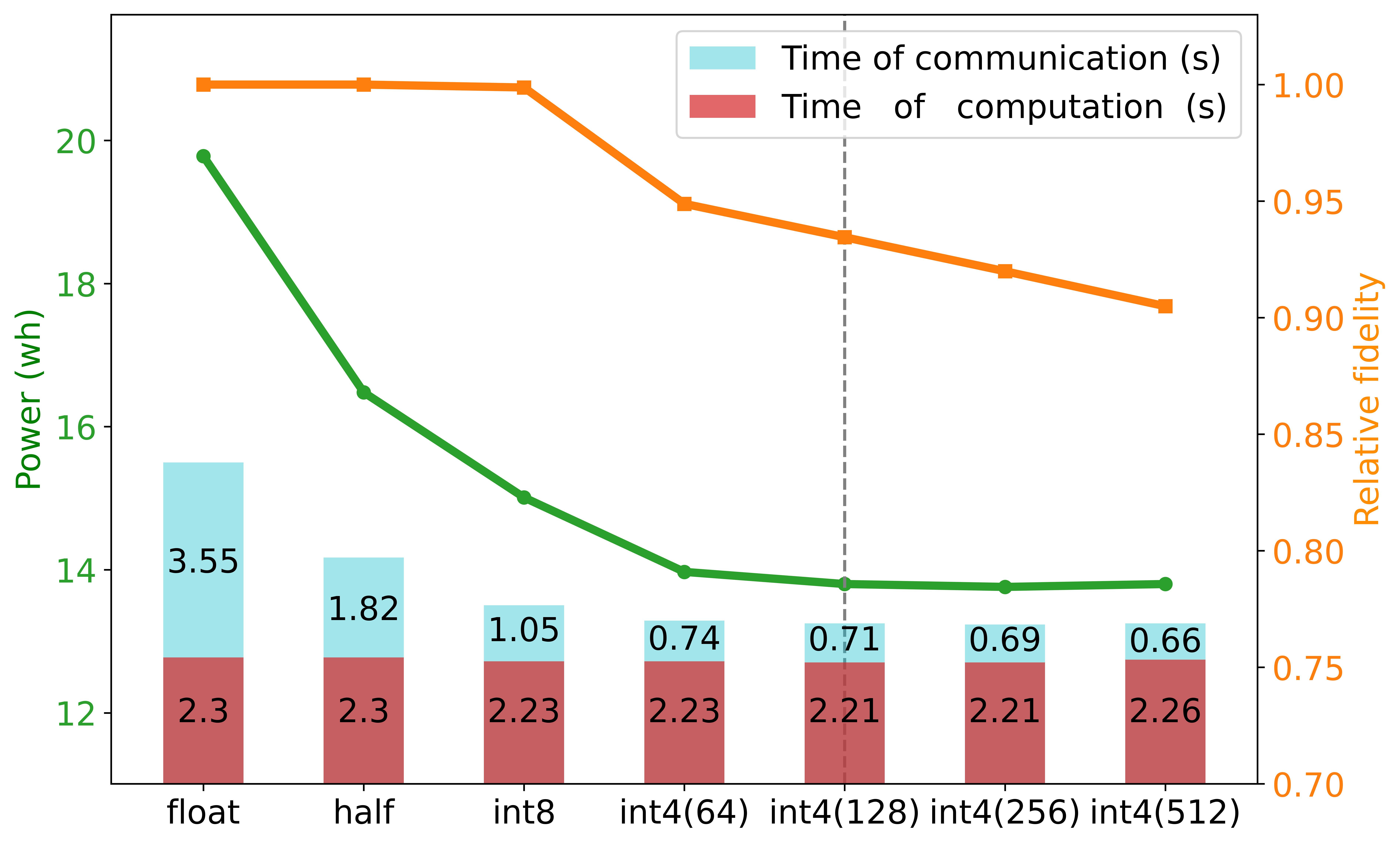}
\caption{Time, energy, and relative fidelity after inter-node quantization on 4T tensor network. The orange curve represents relative fidelity, The green curve represents energy consumption, and the bar chart represents the calculation and communication time. The dashed line represents the experimental plan we ultimately adopted, int4 (128).}
\label{fig:inter_node}
\end{figure}

%half and int8性能 （%campared with complex64）
Despite the negative benefits of quantization during intra-node communication, Equations \ref{eq:time_all2all} and \ref{eq:energy} suggest that quantization can have positive effects on inter-node communication.
To provide a more precise and concrete demonstration, we conducted an end-to-end subtask of a 4T TN. The result is shown in Fig.~\ref{fig:inter_node}.

We observe that the total time and energy consumption gradually decrease from float-precision to int4 (128), but remain relatively constant after int4 (128). The performance improvement from float-precision to int4 (128) is greater than the decrease in relative fidelity. Meanwhile, the improvement after int4 (256) is smaller than the decrease in relative fidelity. In summary, utilizing int4 with 128 being the group size for inter-node quantization yields the maximum positive impact. We adopt this as the final quantization scheme for our task, with a 6.55\% loss in relative fidelity, the time decreased by 50.08\%, and the energy consumption decreased by 30.23\%.

% In terms of inter-node communication quantization, we first used half and Int8 to quantify the global tensor. We found that the time of a single subtask decreased by x\% and 7\%, the energy consumption decreased by x\% and 0.7\%, and the fidelity loss was only 0.03\% and 0.1\%. This prompted us to consider lower precision quantization. We conducted an int4 quantization experiment(see Fig.~\ref{fig:quantize2}). 

% 全性能过度数据（文，表，图）

% %总结全性能文，表，图（%campared with complex64）
% From the image, we can observe that the total time and energy consumption gradually decrease from half to int4(128), and remain almost unchanged after int(256). The performance improvement from half to int(128) is greater than the decrease in relative fidelity, while the performance improvement after int(256) is smaller than the decrease in relative fidelity. In summary, selecting int(128) as the group size for inter-node quantization in int4 can bring the maximum positive benefit, and we will use this as the final quantization scheme for our task. In the end, with a loss of 2\% in relative fidelity, the time decreased by 7.03\%, and the energy consumption decreased by 3.41\%.

\vspace{-0.2cm}
\begin{table*}
\caption{Assessment the Impact of Proposed Methods on the Overall Performance of 4T Tensor Network without post-processing.}
\label{table:melting}
\begin{tabular}{@{}ccccccccc@{}}
\toprule
\multirow{2}{*}{\begin{tabular}[c]{@{}c@{}}Data type of \\ computation\end{tabular}} &
  \multirow{2}{*}{\begin{tabular}[c]{@{}c@{}}Data type of \\ communication\end{tabular}} &
  \multicolumn{3}{c}{Hybrid communication} &
  \multirow{2}{*}{\begin{tabular}[c]{@{}c@{}}Other \\ optimizations\end{tabular}} &
  \multirow{2}{*}{nodes} &
  \multirow{2}{*}{Energy (wh)} &
  \multirow{2}{*}{Fidelity(\%)} \\ \cmidrule(lr){3-5}
      &           & Yes/No? & Inter (GB)/GPU & Intra (GB)/GPU &     &   &       &        \\ \midrule
float & float     & No      & 36         & --             & No  & 8 & 19.78 & 1      \\ \midrule
float & half      & No      & 18             & --             & No  & 8 & 16.48 & 99.999 \\ \midrule
half  & half      & No      & 36        & ---            & No  & 4 & 13.03 & 99.995 \\ \midrule
half  & half      & Yes     & 28         & 20             & No  & 4 & 12.67 & 99.995 \\ \midrule
half  & half      & Yes     & 24             & 40       & Yes & 2 & 10.57 & 99.965 \\ \midrule
half  & int8      & Yes     & 12             & 40       & Yes & 2 & 10.12 & 99.912 \\ \midrule
half  & int4(128) & Yes     & 6              & 40       & Yes & 2 & 9.89  & 98.007 \\ \bottomrule
\end{tabular}
\end{table*}

% \vspace{-0.4cm}
\subsection{Assessment of the proposed techniques}

% Please add the following required packages to your document preamble:
% \usepackage{booktabs}
% \usepackage{multirow}

In this section, we aim to demonstrate how the proposed innovations improve performance incrementally. 
% We assess performance on a cluster of A100 GPUs, with an interconnect bandwidth of 25 GB/s. The subnetwork being evaluated is a 4T tensor contraction, with no top-k method employed.
As shown in Table \ref{table:melting}, each row represents a single subtask. We start from the baseline case, where no optimizations or quantizations are applied, using single-precision floating-point data types and setting the result as a benchmark.

Initially, we decreased the quantity of data by converting single-precision to half-precision for communication, hence leading to a 16.68\% reduction in energy consumption with negligible loss of fidelity.
Additionally, we implement half-precision for computational tasks, reducing the minimum number of nodes needed for a sub-network from 8 to 4. This change leads to a 20.93\% decrease in energy usage while maintaining a negligible loss of accuracy.

Next, we split inter-node communication tasks with intra-node communication, as certain tensor network calculation steps do not require data to be permuted across all nodes. By redirecting some communication burden to intra-node, we observe an additional 2.76\% reduction in energy consumption with minimal fidelity degradation. 
Subsequently, by leveraging the features of the 4T network, we implement a recomputation algorithm that reduces the minimal number of nodes required for a sub-network from 4 to 2, achieving a 16.57\% energy reduction with no significant loss of fidelity. 

Our final experiment employs int8 and int4 quantization for communication, where energy reductions of 4.25\% and 6.43\%, respectively, are achieved with fidelity losses within 2\%. 
In conclusion, the proposed innovations progressively contribute to improving performance, enabling substantial energy savings and maintaining high fidelity in tensor contraction operations.

\vspace{-0.2cm}
\subsection{Verification and simulation of Sycamore circuits}
\label{sec:end2end}
\begin{table*}[tb]
\caption{The metrics and results of the simulated Sycamore experiment. Bold numbers indicate that the result is better than sycamore.}
\label{table:end2end}
\begin{tabular}{ccccc}
\hline
methods &
  \begin{tabular}[c]{@{}c@{}}4T\\ no post-processing\end{tabular} &
  \begin{tabular}[c]{@{}c@{}}4T\\ post-processing\end{tabular} &
  \begin{tabular}[c]{@{}c@{}}32T\\ no post-processing\end{tabular} &
  \begin{tabular}[c]{@{}c@{}}32T\\ post-processing\end{tabular} \\ \hline
Time complexity (FLOP)        & $4.7*10^{17}$       & $7.9*10^{16}$       & $1.3*10^{17}$ & $1.6*10^{16}$ \\ \hline
Memory complexity (elements) &
  $3.1*10^{15}$ &
  $6.4*10^{14}$ &
  $1.3*10^{15}$ &
  $1.6*10^{14}$ \\ \hline
XEB value (\%)               & 0.2036         & 0.2059          & 0.21194                    & 0.2158                     \\ \hline
Efficiency (\%)               & 21.09          & 18.14           & 16.65                      & 17.09                      \\ \hline
\textcolor{red}{Total number of subtasks}     & $2^{18}$ & $2^{18}$ & $2^{12}$ & $2^{12}$ \\ \hline
\textcolor{red}{Number of subtasks conducted}    & 528            & 84              & 9                          & 1                          \\ \hline
Nodes per subtask           & 2            &2            & 32                        & 32                          \\ \hline
Memory/Multi-node level (TB) & 1.25           & 1.25            & 20                         & 20                         \\ \hline

Computer resource (A100)     & 2112           & 96              & 2304                       & 256                        \\ \hline
Time-to-solution (s)         & \textbf{32.51} & \textbf{133.15} & \textbf{14.22}             & \textbf{17.18}             \\ \hline
Energy consumption (kwh)     & 5.77           & \textbf{1.12}   & \textbf{2.39}              & \textbf{0.29}              \\ \hline
\end{tabular}
\end{table*}

% Our three-level scheme can achieve a large number of parallel subtasks to process independent sliced tensors that can be performed on each multi-node level machine. A final reduction will be conducted at the end to collect the results and output the probability.

% To sample from the Sycamore circuit, we choose 10 qubits as open qubits to generate $\tilde{S}$ with $|\tilde{S}| = (3 \times 10^6) \times 2^{10}$ and aim to obtain $k = 3 \times 10^6$ uncorrelated samples via the top-k method.  

In order to sample from the Sycamore circuit, we have applied our techniques to two large-scale tensor networks, namely 4T and 32T, with or without post selection (also known as post processing) \cite{leapfrogging}, aiming for $k = 3 \cdot 10^6$ uncorrelated samples. Our experiment has attained a peak half-precision performance of \textcolor{red}{561} PFLOPS, with approximately 20\% of efficiency achievable through the application of the proposed techniques. A summary of our verification details is presented in Table \ref{table:end2end}, \textcolor{red}{where bold numbers indicate that the metrics performed than Sycamore.} Therefore, all test cases have surpassed the time consumed over Sycamore, and three cases have surpassed energy consumption. Among these, our best case demonstrates one magnitude of order less in both time and energy compared to Sycamore (32T with post-processing).
% 312TFLOPS * 2304(张卡) = 312*10**12 * 2304 = 718.8 PLOPS

We compared the four cases of our simulation for the task of generating
bitstring samples with a bounded fidelity of 0.002. Our analysis focused on three key aspects: the influence of post-selection, the size of the tensor network, and the scalability of the approach.
\vspace{-0.2cm}
\subsubsection{Assessment of Post-processing}As previously mentioned, the extensive tensor network can be divided into self-contained sub-networks, each of which is capable of being contracted at a single multi-node level. In the execution of our simulation, by utilizing a post-selection approach, it is only necessary to perform approximately $11.1\% - 15.9\%$ of the tasks that would have been required without post-selection to achieve an XEB value of 0.002. Thus, comparing the methods with and without post-selection, we can conclude that employing post-selection significantly reduces both time and space complexity.
\vspace{-0.2cm}
\subsubsection{Assessment of size of tensor network}The space complexity of the algorithm is proportional to $s2^M$, where s represents the size of the data type and M corresponds to the targeted contraction treewidth, which can be manually specified in the slicing algorithm. Consequently, the space complexity is completely controlled. Comparing methods between different sizes of tensor networks, it is evident that the time and space complexity decrease as the size of the tensor network increases (at the global level, not for a single multi-node level).

\begin{figure}[h]
\includegraphics[width=\columnwidth]{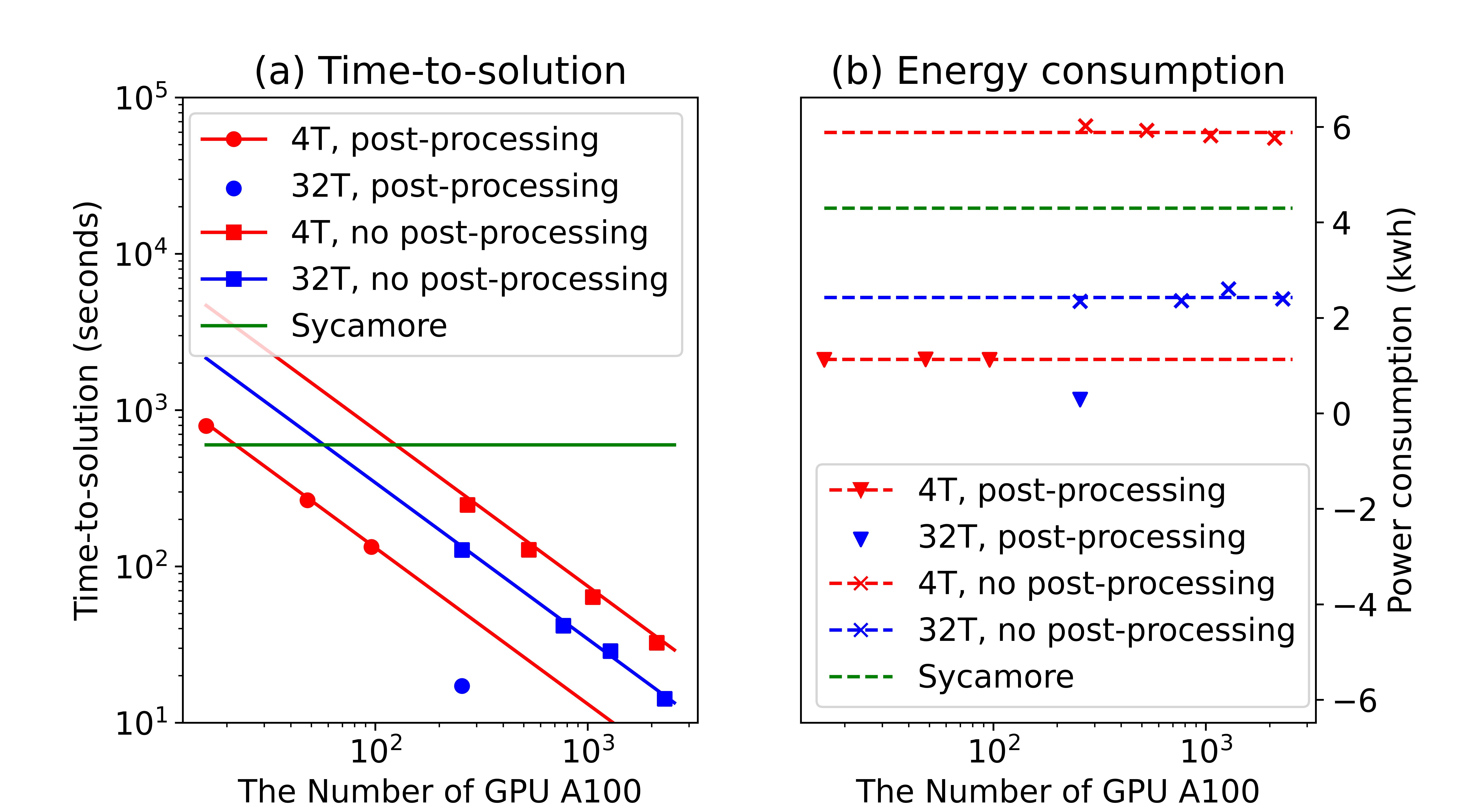}
\caption{The influence of global memory usage on (a) time-to-solution and (b) energy consumption The time-to-solution and energy consumption for simulating a quantum RCS experiment with XEB = 0.002 for samples obtained using different numbers of GPUs. In the 32T tensor network with post-processing, achieving XEB = 0.002 only requires a single multi-node task. Therefore, there is no need for parallelizing subtasks, resulting in just one point (blue circle in (a) and blue triangle in (b)), without any fitting lines.}
\label{fig:time2solution}
\end{figure}

\vspace{-0.2cm}
\subsubsection{Scalibility Exploration}In Fig.~\ref{fig:time2solution}, we observe a linear decay in the total time-to-solution as we increase the number of GPUs used for computation, due to the parallel-friendly feature of the slicing algorithm and three-level parallel scheme. Specifically, our method achieves strong scalability from 128 GPUs (16 nodes) to 768 GPUs (96 nodes) for 4T tensor network with post-selection, from 271 GPUs (34 nodes) to 2112 GPUs (264 nodes) for 4T tensor network without post-selection, from 256 GPUs ( 2 nodes) to 2304 GPUs (288 nodes) for 32T tensor network without post-selection in terms of time. Moreover, the energy consumption remains at a constant level with an increase in nodes.

% We observe an energy consumption of approximately 1.13 kWh for the 4T tensor network and 0.29 kWh for the 32T tensor network, which is less than Sycamore's 1.43 kWh energy consumption.

% where each contraction subtask requires 1.1029 × 1014 FLOPS
% and can be completed in 3.09 seconds on 8 NVIDIA A100 GPUs. Furthermore, we
% observe a linear decay of the total time-to-solution as we increase the number of
% GPUs used in computation (Fig. 7), which clearly demonstrates the parallelism of the
% implementation of our algorithm. In particular, when using 1,432 GPUs, the total timeto-
% solution for sampling 3 × 106 bitstrings is 86.4 sec, which is less than Sycamore’s
% 600-second time-to-solution.

% The results in Section \ref{sec:end2end} also indicate that we can smoothly transition the size of a tensor network, moving from 4T to 32T.

%6 Performance Measurement
%7 Performance Results

%\section{Implications}
\vspace{-0.0cm}
\section{Conclusion}
Previous studies have predominantly focused on identifying optimal paths for classical devices to efficiently compute tensor networks and store tensors within time and memory constraints, from an algorithmic perspective. In contrast, our work aims to enhance the capabilities of multi-node computational clusters and reduce computational complexity. This is achieved through the implementation of a three-level parallel scheme coupled with a carefully crafted inter-node and intra-node hybrid communication approach.

To significantly mitigate overhead costs associated with large tensors, we propose a low-precision communication approach and an einsum extension for complex-half data. Additionally, we introduce several specialized techniques to expedite specific tasks.

% We conducted experiments to evaluate the proposed techniques and determine the optimal parameters for communication and overall strategy. Ultimately, we performed experiments on two large-scale tensor networks, with or without post-selection, to generate 3 million uncorrelated samples with XEB values of $0.002$. This demonstrated scalability in two dimensions: the size of the tensor network and the number of GPUs utilized for executing parallel and independent sub-tasks.

\textcolor{red}{In our best-case without post-processing, we have reduced the time-to-solution to 14.22 seconds, with a power consumption of 2.39 kwh with fidelity of 0.002. Then with the technique of post-processing, we achieve our best performance,} we remarkably reduce the time-to-solution for sampling $3 \times 10^6$ bitstrings to just 17.18 seconds, with a power consumption of 0.29 kWh. This represents a one-order-of-magnitude reduction in both time and energy compared to the Google's quantum processor Sycamore.

In this work, by leveraging state-of-the-art algorithms and a highly optimized scalable tensor network computation system based on GPU clusters, we have significantly surpassed Sycamore in both computing speed and energy consumption. We experimentally \textcolor{red}{challenge} Google's initial claim of quantum supremacy. Our work resets the boundaries of classical computing capabilities and we hope to help the community evaluate quantum advantage more robustly in future quantum computational advantage experiments.

In future work, our techniques supporting large-scale tensor networks can be extended beyond merely RQC sampling simulation. They can be directly applied to diverse fields like quantum computing simulator \cite{guerreschi2020intel}, condensed matter physics \cite{liu2021tropical} and combinatorial optimization \cite{liu2023computing}. This extended application intends to empower the scalability of problem models and facilitate the resolution of realistic challenges such as satisfiability, set packing, clique problems, etc.

% In future work, we anticipate developing low-precision data formats such as Int8 and int4 for tensor network computation, to further minimize input-output time and enhance computational efficiency. Additionally, we plan to design a pipelined scheme to overlap communication latency.
\begin{acks}
This work was supported by the National Key R$\&$D Program of China (No. 2022ZD0160201, No. 2022ZD0160101).
% The corresponding authors of this paper are Han-Sen Zhong (zhonghansen@pjlab.org.cn), Chao-Yang Lu (cylu@ustc.edu.cn), Xingcheng Zhang (zhangxingcheng@pjlab.org.cn) and Zhilin Pei (peizhilin@pjlab.org.cn).
\end{acks}
\vspace{-0.3cm}
\bibliographystyle{unsrt}
\bibliography{./Manuscript}

\begin{thebibliography}{10}

\bibitem{Quantum_supremacy}
Frank Arute, Kunal Arya, Ryan Babbush, Dave Bacon, Joseph~C Bardin, Rami Barends, Rupak Biswas, Sergio Boixo, Fernando~GSL Brandao, David~A Buell, et~al.
\newblock Quantum supremacy using a programmable superconducting processor.
\newblock {\em Nature}, 574(7779):505--510, 2019.

\bibitem{jiuzhang1}
Han-Sen Zhong, Hui Wang, Yu-Hao Deng, Ming-Cheng Chen, Li-Chao Peng, Yi-Han Luo, Jian Qin, Dian Wu, Xing Ding, Yi~Hu, et~al.
\newblock Quantum computational advantage using photons.
\newblock {\em Science}, 370(6523):1460--1463, 2020.

\bibitem{jiuzhang2}
Han-Sen Zhong, Yu-Hao Deng, Jian Qin, Hui Wang, Ming-Cheng Chen, Li-Chao Peng, Yi-Han Luo, Dian Wu, Si-Qiu Gong, Hao Su, et~al.
\newblock Phase-programmable gaussian boson sampling using stimulated squeezed light.
\newblock {\em Physical review letters}, 127(18):180502, 2021.

\bibitem{jiuzhang3}
Yu-Hao Deng, Yi-Chao Gu, Hua-Liang Liu, Si-Qiu Gong, Hao Su, Zhi-Jiong Zhang, Hao-Yang Tang, Meng-Hao Jia, Jia-Min Xu, Ming-Cheng Chen, et~al.
\newblock Gaussian boson sampling with pseudo-photon-number-resolving detectors and quantum computational advantage.
\newblock {\em Physical review letters}, 131(15):150601, 2023.

\bibitem{zuchongzhi56qubit}
Yulin Wu, Wan-Su Bao, Sirui Cao, Fusheng Chen, Ming-Cheng Chen, Xiawei Chen, Tung-Hsun Chung, Hui Deng, Yajie Du, Daojin Fan, et~al.
\newblock Strong quantum computational advantage using a superconducting quantum processor.
\newblock {\em Physical review letters}, 127(18):180501, 2021.

\bibitem{zuchongzhi60qubit}
Qingling Zhu, Sirui Cao, Fusheng Chen, Ming-Cheng Chen, Xiawei Chen, Tung-Hsun Chung, Hui Deng, Yajie Du, Daojin Fan, Ming Gong, et~al.
\newblock Quantum computational advantage via 60-qubit 24-cycle random circuit sampling.
\newblock {\em Science bulletin}, 67(3):240--245, 2022.

\bibitem{bertels2021quantum}
Koen Bertels, Aritra Sarkar, and Imran Ashraf.
\newblock Quantum computing—from nisq to pisq.
\newblock {\em IEEE Micro}, 41(5):24--32, 2021.

\bibitem{zlokapa2023boundaries}
Alexander Zlokapa, Benjamin Villalonga, Sergio Boixo, and Daniel~A Lidar.
\newblock Boundaries of quantum supremacy via random circuit sampling.
\newblock {\em npj Quantum Information}, 9(1):36, 2023.

\bibitem{villalonga2020establishing}
Benjamin Villalonga, Dmitry Lyakh, Sergio Boixo, Hartmut Neven, Travis~S Humble, Rupak Biswas, Eleanor~G Rieffel, Alan Ho, and Salvatore Mandr{\`a}.
\newblock Establishing the quantum supremacy frontier with a 281 pflop/s simulation.
\newblock {\em Quantum Science and Technology}, 5(3):034003, 2020.

\bibitem{haner20175}
Thomas H{\"a}ner and Damian~S Steiger.
\newblock 0.5 petabyte simulation of a 45-qubit quantum circuit.
\newblock In {\em Proceedings of the International Conference for High Performance Computing, Networking, Storage and Analysis}, pages 1--10, 2017.

\bibitem{wu2019full}
Xin-Chuan Wu, Sheng Di, Emma~Maitreyee Dasgupta, Franck Cappello, Hal Finkel, Yuri Alexeev, and Frederic~T Chong.
\newblock Full-state quantum circuit simulation by using data compression.
\newblock In {\em Proceedings of the International Conference for High Performance Computing, Networking, Storage and Analysis}, pages 1--24, 2019.

\bibitem{Alibaba_19days}
Cupjin Huang, Fang Zhang, Michael Newman, Junjie Cai, Xun Gao, Zhengxiong Tian, Junyin Wu, Haihong Xu, Huanjun Yu, Bo~Yuan, et~al.
\newblock Classical simulation of quantum supremacy circuits.
\newblock {\em arXiv preprint arXiv:2005.06787}, 2020.

\bibitem{Sunway_304s}
Yong Liu, Xin Liu, Fang Li, Haohuan Fu, Yuling Yang, Jiawei Song, Pengpeng Zhao, Zhen Wang, Dajia Peng, Huarong Chen, et~al.
\newblock Closing the "quantum supremacy" gap: achieving real-time simulation of a random quantum circuit using a new sunway supercomputer.
\newblock In {\em Proceedings of the International Conference for High Performance Computing, Networking, Storage and Analysis}, pages 1--12, 2021.

\bibitem{liu2022validating}
Yong Liu, Yaojian Chen, Chu Guo, Jiawei Song, Xinmin Shi, Lin Gan, Wenzhao Wu, Wei Wu, Haohuan Fu, Xin Liu, et~al.
\newblock Validating quantum-supremacy experiments with exact and fast tensor network contraction.
\newblock {\em arXiv preprint arXiv:2212.04749}, 2022.

\bibitem{guo2019general}
Chu Guo, Yong Liu, Min Xiong, Shichuan Xue, Xiang Fu, Anqi Huang, Xiaogang Qiang, Ping Xu, Junhua Liu, Shenggen Zheng, et~al.
\newblock General-purpose quantum circuit simulator with projected entangled-pair states and the quantum supremacy frontier.
\newblock {\em Physical review letters}, 123(19):190501, 2019.

\bibitem{li2019quantum}
Riling Li, Bujiao Wu, Mingsheng Ying, Xiaoming Sun, and Guangwen Yang.
\newblock Quantum supremacy circuit simulation on sunway taihulight.
\newblock {\em IEEE Transactions on Parallel and Distributed Systems}, 31(4):805--816, 2019.

\bibitem{512GPUs_15h}
Feng Pan, Keyang Chen, and Pan Zhang.
\newblock Solving the sampling problem of the sycamore quantum circuits.
\newblock {\em Phys. Rev. Lett.}, 129:090502, Aug 2022.

\bibitem{60GPUs_5days}
Feng Pan and Pan Zhang.
\newblock Simulation of quantum circuits using the big-batch tensor network method.
\newblock {\em Phys. Rev. Lett.}, 128:030501, Jan 2022.

\bibitem{leapfrogging}
Xian-He Zhao, Han-Sen Zhong, Feng Pan, Zi-Han Chen, Rong Fu, Zhongling Su, Xiaotong Xie, Chaoxing Zhao, Pan Zhang, Wanli Ouyang, et~al.
\newblock Leapfrogging sycamore: Harnessing 1432 gpus for 7$\times$ faster quantum random circuit sampling.
\newblock {\em arXiv preprint arXiv:2406.18889}, 2024.

\bibitem{post-process}
Xun Gao, Marcin Kalinowski, Chi-Ning Chou, Mikhail~D. Lukin, Boaz Barak, and Soonwon Choi.
\newblock Limitations of linear cross-entropy as a measure for quantum advantage.
\newblock {\em PRX Quantum}, 5:010334, Feb 2024.

\bibitem{Lifetime-Based}
Yaojian Chen, Yong Liu, Xinmin Shi, Jiawei Song, Xin Liu, Lin Gan, Chu Guo, Haohuan Fu, Jie Gao, Dexun Chen, and Guangwen Yang.
\newblock Lifetime-based optimization for simulating quantum circuits on a new sunway supercomputer.
\newblock In {\em Proceedings of the 28th ACM SIGPLAN Annual Symposium on Principles and Practice of Parallel Programming}, PPoPP '23, page 148–159, New York, NY, USA, 2023. Association for Computing Machinery.

\bibitem{brandhofer2023optimal}
Sebastian Brandhofer, Ilia Polian, and Kevin Krsulich.
\newblock Optimal partitioning of quantum circuits using gate cuts and wire cuts.
\newblock {\em IEEE Transactions on Quantum Engineering}, 2023.

\bibitem{pan2023efficient}
Feng Pan, Hanfeng Gu, Lvlin Kuang, Bing Liu, and Pan Zhang.
\newblock Efficient quantum circuit simulation by tensor network methods on modern gpus.
\newblock {\em arXiv preprint arXiv:2310.03978}, 2023.

\bibitem{de2019massively}
Hans De~Raedt, Fengping Jin, Dennis Willsch, Madita Willsch, Naoki Yoshioka, Nobuyasu Ito, Shengjun Yuan, and Kristel Michielsen.
\newblock Massively parallel quantum computer simulator, eleven years later.
\newblock {\em Computer Physics Communications}, 237:47--61, 2019.

\bibitem{vidal2003efficient}
Guifr{\'e} Vidal.
\newblock Efficient classical simulation of slightly entangled quantum computations.
\newblock {\em Physical review letters}, 91(14):147902, 2003.

\bibitem{state_vector}
Koen De~Raedt, Kristel Michielsen, Hans De~Raedt, Binh Trieu, Guido Arnold, Marcus Richter, Th~Lippert, Hiroshi Watanabe, and Nobuyasu Ito.
\newblock Massively parallel quantum computer simulator.
\newblock {\em Computer Physics Communications}, 176(2):121--136, 2007.

\bibitem{villalonga2019flexible}
Benjamin Villalonga, Sergio Boixo, Bron Nelson, Christopher Henze, Eleanor Rieffel, Rupak Biswas, and Salvatore Mandr{\`a}.
\newblock A flexible high-performance simulator for verifying and benchmarking quantum circuits implemented on real hardware.
\newblock {\em npj Quantum Information}, 5(1):86, 2019.

\bibitem{chen2018classical}
Jianxin Chen, Fang Zhang, Cupjin Huang, Michael Newman, and Yaoyun Shi.
\newblock Classical simulation of intermediate-size quantum circuits.
\newblock {\em arXiv preprint arXiv:1805.01450}, 2018.

\bibitem{cotengra}
Johnnie Gray and Stefanos Kourtis.
\newblock Hyper-optimized tensor network contraction.
\newblock {\em Quantum}, 5:410, 2021.

\bibitem{liu2021redefining}
Xin Liu, Chu Guo, Yong Liu, Yuling Yang, Jiawei Song, Jie Gao, Zhen Wang, Wenzhao Wu, Dajia Peng, Pengpeng Zhao, et~al.
\newblock Redefining the quantum supremacy baseline with a new generation sunway supercomputer.
\newblock {\em arXiv preprint arXiv:2111.01066}, 2021.

\bibitem{Vectorquantization}
Ze-bin Wu and Jun-qing Yu.
\newblock Vector quantization: a review.
\newblock {\em Frontiers of Information Technology \& Electronic Engineering}, 20(4):507--524, 2019.

\bibitem{GDRQ}
Haibao Yu, Tuopu Wen, Guangliang Cheng, Jiankai Sun, Qi~Han, and Jianping Shi.
\newblock Gdrq: Group-based distribution reshaping for quantization.
\newblock {\em arXiv preprint arXiv:1908.01477}, 2019.

\bibitem{nvml}
Nvidia management library (nvml).
\newblock \url{https://developer.nvidia.com/nvidia-management-library-nvml}.
\newblock Accessed: 2024-03-27.

\bibitem{guerreschi2020intel}
Gian~Giacomo Guerreschi, Justin Hogaboam, Fabio Baruffa, and Nicolas~PD Sawaya.
\newblock Intel quantum simulator: A cloud-ready high-performance simulator of quantum circuits.
\newblock {\em Quantum Science and Technology}, 5(3):034007, 2020.

\bibitem{liu2021tropical}
Jin-Guo Liu, Lei Wang, and Pan Zhang.
\newblock Tropical tensor network for ground states of spin glasses.
\newblock {\em Physical Review Letters}, 126(9):090506, 2021.

\bibitem{liu2023computing}
Jin-Guo Liu, Xun Gao, Madelyn Cain, Mikhail~D Lukin, and Sheng-Tao Wang.
\newblock Computing solution space properties of combinatorial optimization problems via generic tensor networks.
\newblock {\em SIAM Journal on Scientific Computing}, 45(3):A1239--A1270, 2023.

\end{thebibliography}

\end{document}